\documentclass[useAMS,usenatbib]{mn2e}

\usepackage[pdftex]{graphicx}

\usepackage{relsize}

\newcommand{\hi}{%
  \relax
  \ifmmode
    \textrm{\textsc{HI}}
  \else
    \textsc{H{\smaller}I}
  \fi
}

\newcommand{\simgt}{\lower.5ex\hbox{$\; \buildrel > \over \sim \;$}}
\newcommand{\simlt}{\lower.5ex\hbox{$\; \buildrel < \over \sim \;$}}

\title[The Shape of Dark Matter Haloes III. Kinematics and Structure of the \hi Disc]{The Shape of Dark Matter Haloes\\
III. Kinematics and Structure of the \hi disc}
\author[S. P. C. Peters et al.]{S. P. C. Peters$^{1}$,
P. C. van der Kruit$^{1}$\thanks{For more information, 
please contact P.C. van der Kruit at email vdkruit@astro.rug.nl.}, 
R. J. Allen$^{2}$ and K. C. Freeman$^{3}$\\
$^{1}$Kapteyn Astronomical Institute, University of Groningen, P.O.Box 800, 9700AV Groningen, the Netherlands\\
$^{2}$Space Telescope Science Institute, 3700 San Martin Drive, Baltimore, MD 21218, USA\\
$^{3}$Research School of Astronomy and Astrophysics The Australian National University, Cotter Road Weston Creek, ACT 2611,\\
Australia}

\begin{document}

\date{Accepted 2015 month xx. Received 2015 Month xx; in original form 2015 Month xx}
\pagerange{\pageref{firstpage}--\pageref{lastpage}} \pubyear{2015}

\maketitle

\label{firstpage}

\begin{abstract}
We present a new strategy for fitting the structure and kinematics of the \hi 
in edge-on 
                 galaxies using a fit to the terminal-velocity channel maps 
of a  \hi data cube. 
				The strategy can deal with self-absorbing 
\hi gas and the presence of warps.
                The method is first tested on a series of models.
                We demonstrate that fitting optically thin models to real 
galaxies will lead to an overestimation of the thickness and velocity 
                 dispersion, and to a serious underestimation of the \hi 
face-on column densities.
                We subsequently fit both self-absorption and optically 
thin models to the \hi data of six edge-on galaxies. In three of 
                 these we have also measured the velocity dispersion. 
				On average $27\pm6\%$ of the total \hi 
mass of edge-on galaxies is hidden by self-absorption.
				This implies that the \hi mass, thickness 
and velocity dispersion of galaxies is typically underestimated in the 
literature.
\end{abstract}


\section{Introduction}
In Paper I in this series, 
we have presented the \hi observations for eight edge-on late-type galaxies.
One of our main conclusions was that the \hi showed clear signs of 
self-absorption, which created the risk of a rather drastic underestimation 
of the baryonic content of the galaxy.
We developed a new \hi modelling and fitting system called \textsc{Galactus} 
in Paper II to address this problem.
We showed that the \hi self-absorption indeed has a drastic impact on the 
observed maximum surface brightness temperature.
Self-absorption needs to be taken into account when modelling the \hi in 
edge-on galaxies.

The main goal of our project throughout this series of papers is to 
measure the hydrostatics at the central plane of edge-on galaxies.
To get to this, we will need to  derive the rotation curve $v_\textrm{rot}(R)$, 
face-on surface density $A_\hi(R)$, thickness of the HI layer $z_0(R)$ and 
its velocity 
dispersion $\sigma(R)$ accurately.
Secondary parameters, such as the exact kinematic position in RA, DEC and the 
systemic velocity $v_\textrm{sys}$, will also need to be derived.
While warps are of scientific interest, they represent a disturbance of the 
central plane of the galaxy. 
The physics behind warps is not well understood \citep{kf11}.
Using the hydrostatics from the warped region would be a dangerous and 
ultimately futile exercise. 
We therefore refrain from modelling the warps. 
The channel maps of the observations, with the model superimposed on it, 
will be used to estimate the position at which the warp sets in.
Beyond this position, we will not calculate the hydrostatics.

Fitting the \hi structure and kinematics of edge-on galaxies by modelling the 
position-velocity (XV) diagram has a long-standing record of accomplishment.
Various methods have been devised, initially only to derive the rotation curve 
and the
radial distribution of HI surface brightness, and in some cases also the 
flaring (the increasing
thickness as a function of galactocentric radius) of the \hi layer  
\citep[e.g.][and references therein]{Sancisi1979A, vdk81c,   
Garcia-Ruiz2002A, Takamiya2002A, 
Uson2003A, kk04, Kregel2004B}.
More recently, \citet{Olling1996A, Olling1996B} and 
\citet{OBrien2010A,OBrien2010B,OBrien2010C,OBrien2010D} have  measured the 
\hi velocity dispersion. 
The paper by \citet{OBrien2010B} provides a detailed description of the 
various methods and a discussion on the relative merits and pitfalls.

In this paper, we develop a new approach to modelling the \hi properties 
of an edge-on galaxy, which consists of fitting channel maps near the terminal 
velocity of the galaxy.
In Section \ref{sec:HIfittingstrategy}, we explain this fitting strategy in 
more detail.
Section \ref{sec:HIfittingtestmodels} then applies this strategy on a series 
of test models and demonstrates the validity of the method.
The kinematics and structure of six edge-on galaxies is subsequently presented 
in Section \ref{sec:HIfittingresults}.

\section{Fitting Strategy for Edge-on Galaxies}\label{sec:HIfittingstrategy}
In Paper II, we demonstrated that the face-on galaxy NGC\,2403 has at least 10\% 
more \hi mass than an optically thin model would have measured.
Rotating NGC\,2403 to an edge-on geometry and assuming
a spin temperature of 100\,K, 
we found that 22\% of the total \hi mass would have been hidden by 
self-absorption. 
A lower median spin temperature of 80\,K increases this fraction to 30\% 
(see Section 7 of Paper II).
Any attempt to model an edge-on galaxy as optically thin will thus suffer 
from a drastic underestimation of total \hi mass.

While this is troubling enough by itself, other questions regarding the 
effect of the self-absorption can also be raised:
\begin{itemize}
 \item What is the effect of self-absorption on the rotation-curve estimate?
 \item Is the thickness of the disc still measured properly?
 \item What is the impact on the measured velocity dispersion?
\end{itemize}

A common practice for measuring the kinematics of edge-on galaxies is 
through the outer envelope of the (integrated) XV diagram, either adopting 
the peak flux as the rotation at the line of nodes, or by fitting a 
half-Gaussian to the terminal velocities 
\citep[e.g.][and references therein]{Sancisi1979A,Olling1996A,kk04, OBrien2010B, OBrien2010C}.
Both the velocity dispersion and thickness of the disc can 
change the mid-plane surface brightness in a particular channel.
This creates a degeneracy between the two parameters. 
As such, there is no practical way to correct for the 
self-absorption using only the mid-plane or integrated XV-diagrams 
of a galaxy. 
We therefore opt for a different strategy and model 
the channel-maps of the  \hi data cube directly.
An advantage to this strategy is that we can fit all 
four main parameters (i.e. the rotation curve, face-on 
surface density, velocity dispersion and the thickness) 
simultaneously and self-consistently.
After much trial-and-error, we have devised the following 
three-pass strategy for fitting edge-on galaxies, using 
the \textsc{Galactus} fitting code developed in Paper II.

In the first pass, we fix the velocity dispersion at a 
constant value of 10 km/s and fit the rotation curve, 
face-on surface density and thickness of the disc.
The central position ($x$, $y$, $v_\textrm{sys}$) along 
with the position angle $PA$ are also fitted during this pass.
Tests have shown that measuring the exact inclination is 
difficult, so we fix the inclination at $90^\circ$.
We fit the entire \hi data cube. The galaxy has already been projected 
such that the major axis aligns with the horizontal axis as already discussed 
in Paper I.
 
The results are subsequently inspected and corrections made where necessary.
We next fix the position and position angle.
The galaxy is split in two halves, which will be fit separately.
Based on the results from the initial fit, we create a mask, such that 
the terminal velocity is still visible at all radii, but the radii 
at lower rotational velocities are masked. 
The line-of-nodes velocity is expected near the terminal velocity.
We do this because warps and asymmetries, together with the 
self-absorption, lead to degeneracies in the solution and 
thus can confuse the fitting algorithm.
The masking is done by hand. Results from subsequent fits 
are inspected to ensure the line of nodes rotation is well 
beyond the masked positions.
This is effectively creating an outer-envelope mask. Rather 
than just mask the XV diagram, we are now masking the \hi data cube itself.
After both sides have their  \hi data cubes masked, we fit the 
rotation curve, thickness, face-on surface density and velocity dispersion.
The results are again inspected, corrected and refit where necessary.

\begin{figure}
   \centering
   \includegraphics[width=0.48\textwidth]{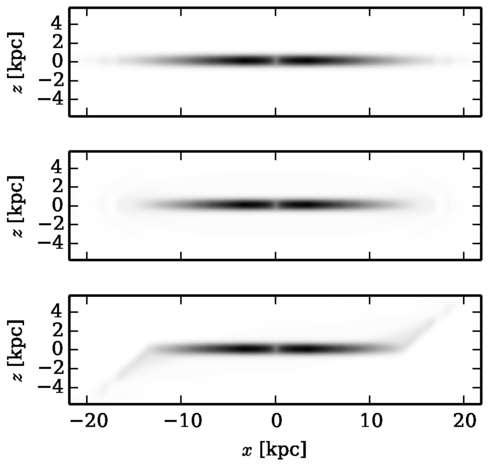}
   \caption[Moment-zero maps of warps]{Moment-zero maps of a model without warp (top), face-on warp (middle) and a side-on warp (bottom).}
   \label{fig:maskingtheenvelope1}
\end{figure}

\begin{figure*}
   \centering
   \includegraphics[width=0.45\textwidth]{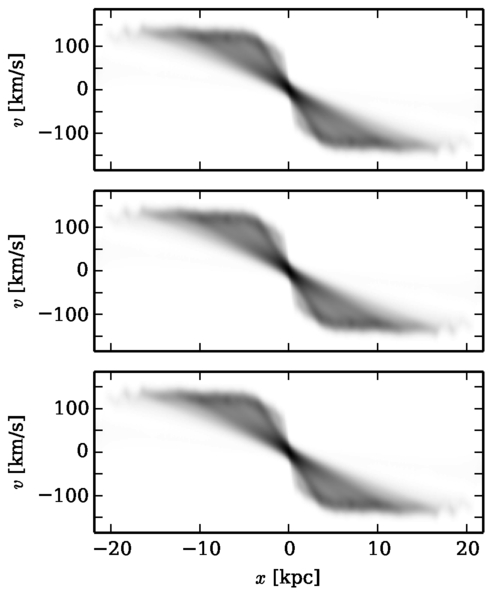}
   \includegraphics[width=0.45\textwidth]{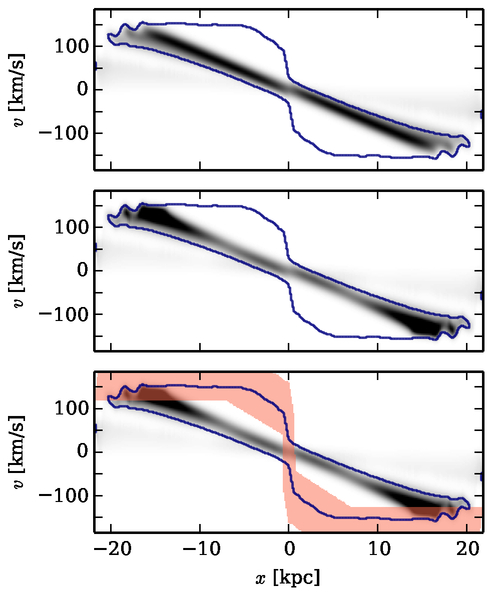}
   \caption[Integrated XV diagram and the effects due to warps]{Left: Integrated XV-diagrams for a model without a warp (top), a line-of-sight warp (middle) and a side-on warp (bottom). Right: Integrated XV diagram of the difference  \hi data cube between the non-warped and face-on warp model (top), and the non-warped and side-on warp model (middle). The blue contours show the outline of the full XV diagram. The bottom panel reproduces the middle panel, but has the envelope mask superimposed on top of it. Note that the fitting occurs on the channel maps rather than the XV-diagram.}
   \label{fig:maskingtheenvelope2}
\end{figure*}

To illustrate the need for an outer-envelope mask, we use 
\textsc{Galactus} to model various types of warps.
We begin with the parameters of the model from Section 
7 of Paper II. This model is based on the \hi in NGC\,2403, as
determined by us using a fitting of the data taking self-absorption 
into account. The models will be simulated 
as perfectly edge-on.
We generate three models: The first will have no warp, while 
the second will have a strong line-of-sight warp (the 
maximum deviation from the mid-plane occurs along the line 
of sight) and the third one a strong side-on warp (the maximum 
deviation from the mid-plane occurs 
perpendicular to the line of sight).
The models are run in self-absorption mode at a spin 
temperature of 100\,K.
Both warps begin at a radius of 13.6\,kpc from the centre of the galaxy.
The offset of the warp above the plane the increases 
linearly with one kpc height for every one kpc radius.
It peaks at a radius of 20.4\,kpc at a height of 6.8\,kpc.
In Figure \ref{fig:maskingtheenvelope1}, we demonstrate 
the zeroth-moment maps for the three models.
The presence of a line-of-sight warp is very hard to 
detect from this image, as is typical for these types 
of warps \citep[e.g.][and references therein]{Gentile2003A}. 
In contrast, the side-on warp is clearly visible.

Shown in Figure \ref{fig:maskingtheenvelope2} are 
the height-integrated XV diagrams of the three models.
While some variation is present between the three 
images, the differences are minor.
To illustrate the effect of the warp, we have created 
two difference XV-maps in Figure \ref{fig:maskingtheenvelope2}.
Here we have subtracted the non-warped  \hi data cube from both warped 
 \hi data cubes and 
integrated the absolute values of these differences along the minor axis.
Both warps create a different signature in these difference maps. 
The line-of-sight warp creates a bar spread evenly over all lower 
velocity channels, while the side-on warp is far more pronounced at larger
positions of $x$.
Unsurprisingly, beyond $x>13.6$\,kpc the warps affect the 
terminal-velocity channels as well and will thus affect the parameters 
extracted from this region.
However, as demonstrated by the superimposed mask in the last panel, 
the outer envelope of the  \hi data cube at radii less than 
13.6 kpc is not affected by the warp.

Using an envelope mask of the  \hi data cube, we can thus measure the parameters 
of an edge-on galaxy inside the warped region, regardless of the 
presence of either a line-of-sight or side-on warp. 
Had we not masked the lower velocities, then the effect of the 
warps would have `confused' the fitting algorithm.

In the final pass, the results are sampled with the Monte-Carlo Markov-Chain 
(MCMC) code \textsc{emcee}, which is used to sample the likelihood 
distribution in each parameter (see Section 2.8 of Paper II 
for more details).
The results, auto-correlation and traces are inspected after a sufficient 
number of samples (i.e. 100.000+) have been drawn.
When the MCMC has settled into a stable distribution, the final 10.000 
samples are used to calculate the parameter distributions.
We visualize these parameter distribution in subsequent figures based 
on the central 68.3, 95.5 and 99.7\% fractions.

We fit each galaxy in two ways: using both an optically thin model and 
a self-absorption model with a spin temperature of 100\,K.
As the physics of warps is not well understood\footnote{see \citet{kf11} 
and references therein}, we do not concern ourselves with warps.
We fit the entire galaxy and determine by eye at what radius the warp 
starts to affect the flaring measurement.
Beyond that radius, the data is considered unreliable.

\section{Testing the Strategy}\label{sec:HIfittingtestmodels}
\begin{figure*}
\centering
\includegraphics[width=0.32\textwidth]{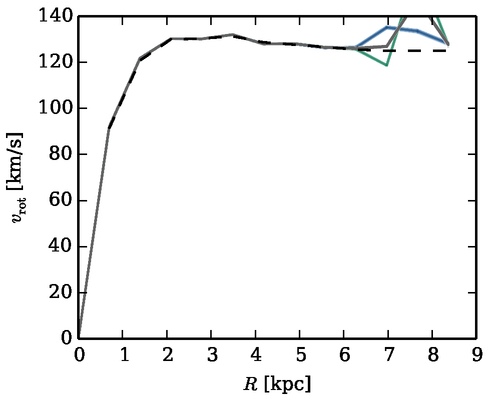}
\includegraphics[width=0.32\textwidth]{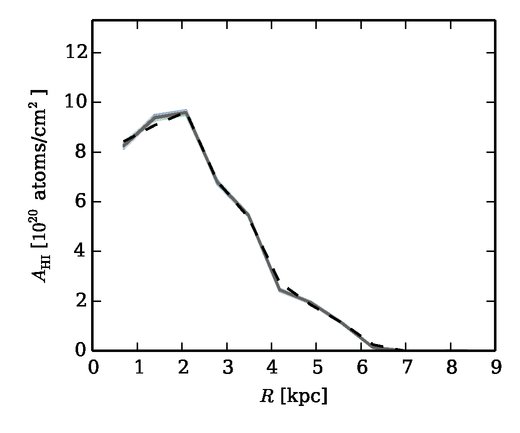}

\includegraphics[width=0.32\textwidth]{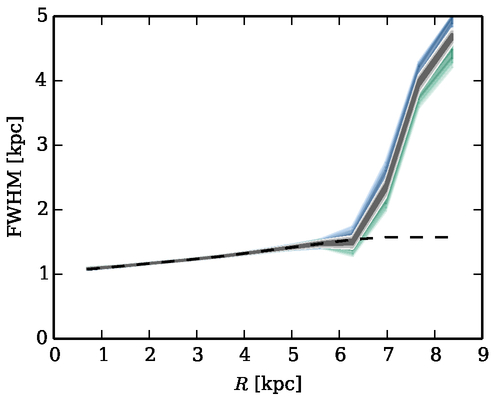}
\includegraphics[width=0.32\textwidth]{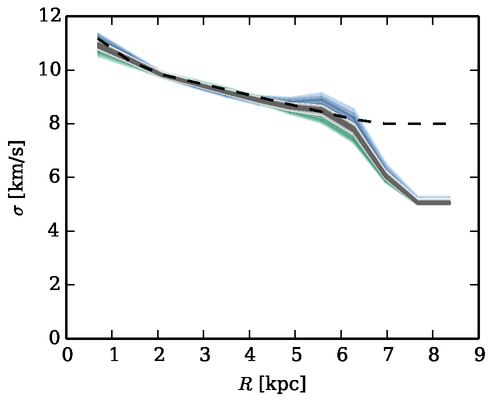}
\caption[Optically thin test model fit (0.1\,K)]{Results from a fit to both sides of an optically thin  test model with low noize (0.1\,K). Top-left panel shows the rotation curve $v_\textrm{rot}$, top-right panel the face-on surface density $A_\hi$. The lower-left panel shows the FWHM thickness of the disc. The velocity dispersion $\sigma$ is shown in the lower-right panel. Green colours denote the left side of the galaxy, blue colours show the right side. The grey band is the combined result. Dotted lines show the true parameters of the model.}\label{fig:testmodel00}
\end{figure*}
\subsection{High S/N optically thin fit}
We first test our strategy on a series of models.
We begin with an optically thin model, with a very low noise at 
$\sigma=0.1$\,K. The model resembles UGC\,7321, except that the 
linear scale is twice as small.
The data is modelled at a distance of 10\,Mpc, with pixels of 
6.8'', a channel width of 3.3\,km/s and a FWHM beam of 13.8''.
At only 48\,K, the maximum surface brightness in the model is 
still low, but the maximum signal to noise is very high at 480.
We present the results of the fit to this model in Figure 
\ref{fig:testmodel00}. 
As is clear from the image, the fit matches the initial near 
perfectly. Only at very large radii, where the \hi surface density is 
lower than about  $10^{20}$ atoms/cm$^{2}$, do we find a sudden increase in FWHM
and decrease in $\sigma$ that are not present in the model. For such levels,
where nearly no \hi is present the fits become unreliable. 
We do find such behaviour in noisier models and in actual fits on our data 
below, usually also at levels below $10^{20}$ atoms/cm$^{2}$. Obviously 
such results should not and will not be used in any further modelling.

\subsection{Optically Thin Fits}
\begin{figure*}
\centering
\includegraphics[width=0.32\textwidth]{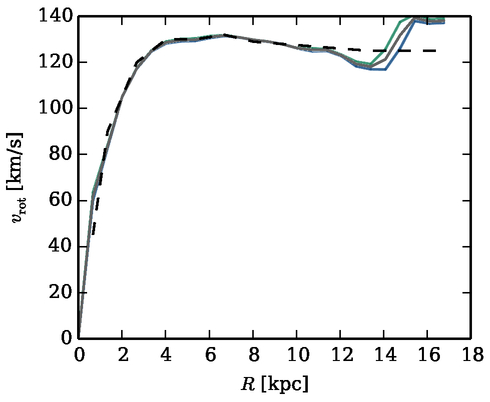}
\includegraphics[width=0.32\textwidth]{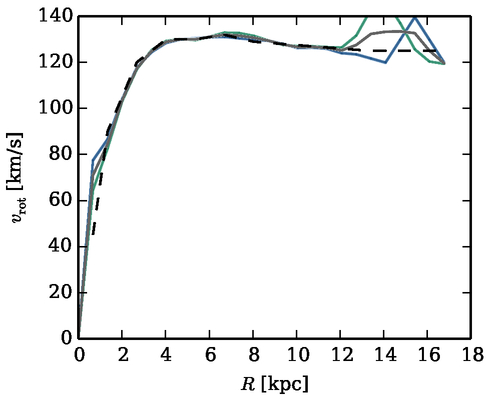}
\includegraphics[width=0.32\textwidth]{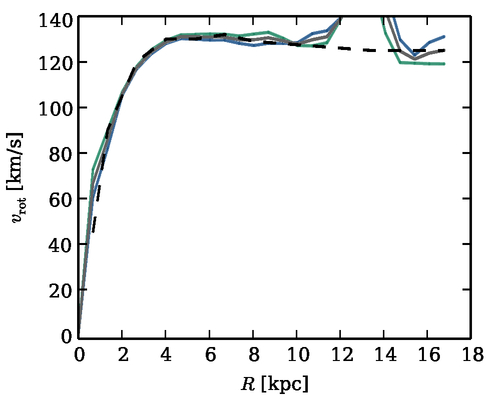}
\includegraphics[width=0.32\textwidth]{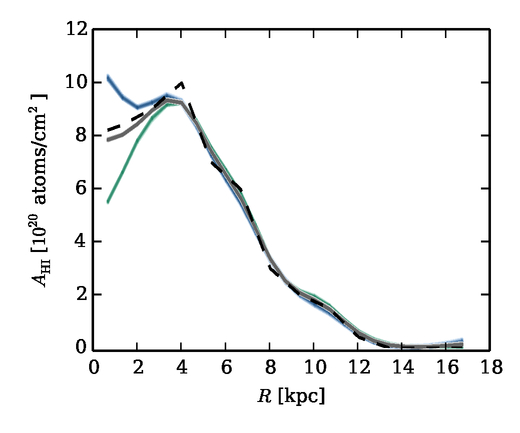}
\includegraphics[width=0.32\textwidth]{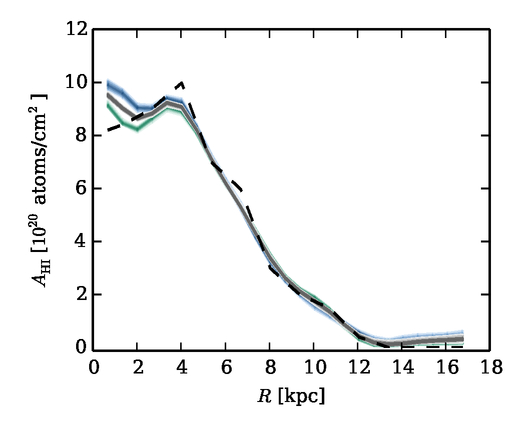}
\includegraphics[width=0.32\textwidth]{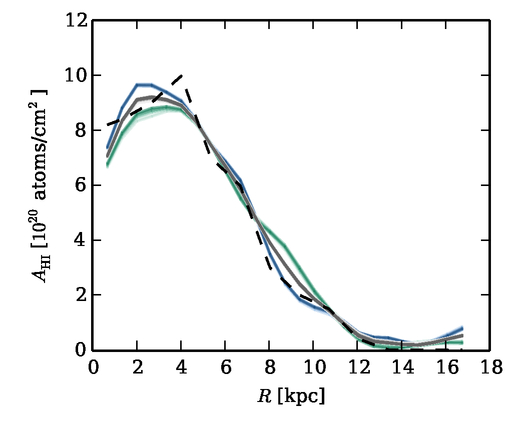}
\includegraphics[width=0.32\textwidth]{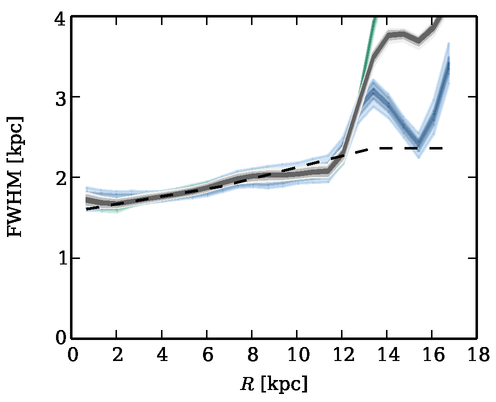}
\includegraphics[width=0.32\textwidth]{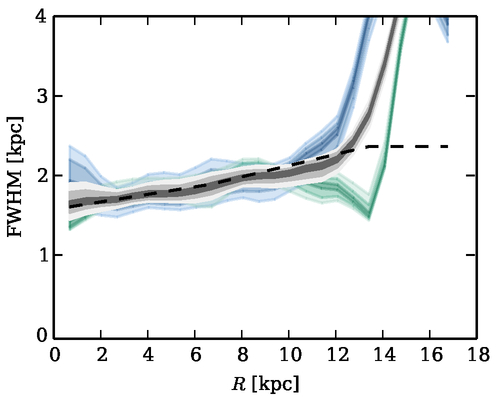}
\includegraphics[width=0.32\textwidth]{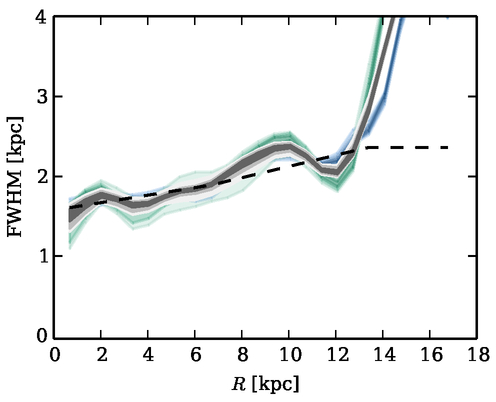}
\includegraphics[width=0.32\textwidth]{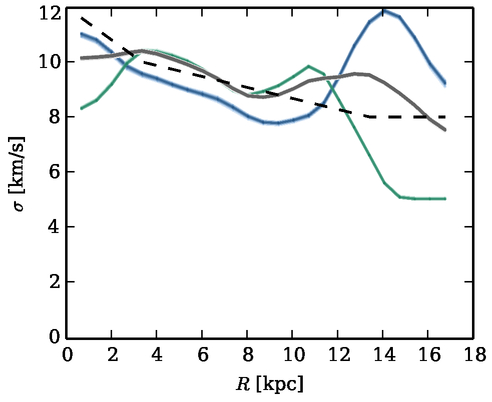}
\includegraphics[width=0.32\textwidth]{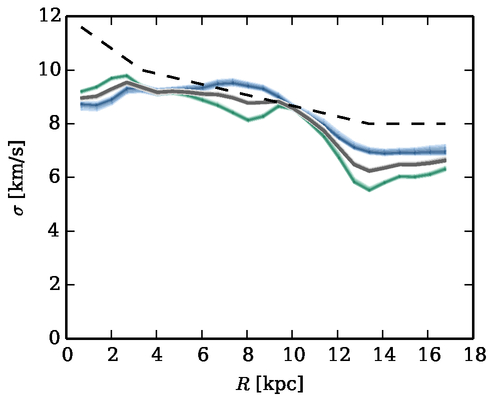}
\includegraphics[width=0.32\textwidth]{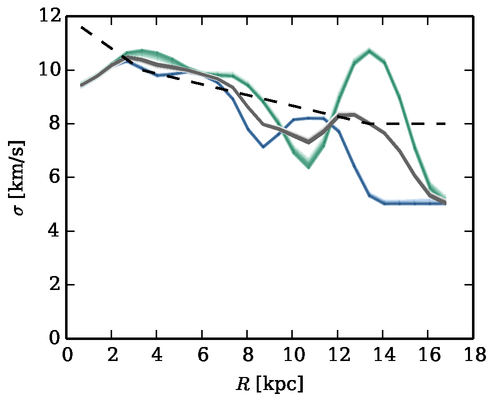}
\caption[Optically thin test model fits (1\,K, 3\,K and 5\,K)]{Optically thin test models with noises of 1\,K, 3\,K and 5\,K (left-to-right). Top panels shows the rotation curve $v_\textrm{rot}$, second row shows the face-on surface density $A_\hi$. Third row shows the thickness of the model. Fourth row shows the fit to the velocity dispersion.  Green colours denote the left side of the galaxy, blue colours show the right side. The grey band is the combined result. Dashed lines show the true parameters of the model.}\label{fig:testmodels-thin}
\end{figure*}

So how does the fitting deteriorate when the noise becomes higher?
We double the size of the model from the previous test, such that the 
radial distance of each value of the parameters is twice as far out. 
The cell size is also doubled to 13.8''. The model now has the same 
linear size as UGC\,7321, on which it is based.
This has the effect of raising the maximum surface brightness 
temperature to a more realistic value $\sim90$\,K.
The models are run with noise levels of 1, 3 and 5\,K.
The maximum signal to noise then becomes 93, 30 and 20. 
The parameters found in the fits on either sides are smoothed.
To create the combined parameters, the unsmoothed fits to either side are 
averaged together and the combined results smoothed.
For the rotation curve, flaring and face-on surface density with a 
kernel of [1/4, 1/2, 1/2] and the velocity dispersion with a kernel of 
[1/3, 1/3, 1/3], as these parameters are found to be more sensitive to noise.
For the combined result, both sides of the galaxy are averaged together 
from the unsmoothed fits, and only then is the combined data smoothed.
We show the results for this fit in Figure \ref{fig:testmodels-thin}.
The rotation curve and the face-on surface density are recovered well 
in all cases.
Measuring the thickness of the disc works reasonably well for the 1 
and 3\,K models -- at least out to 10-12 kpc --, 
but the 5\,K result starts to deviate more.
The velocity dispersion is the hardest parameter to fit. 
Only for the 1\,K model is the input model recovered sufficiently 
well out to 10 kpc, the 3\,K result is marginal and the 5\,K result is 
doubtful even at 8 kpc.

\begin{figure*}
\centering
\includegraphics[width=0.32\textwidth]{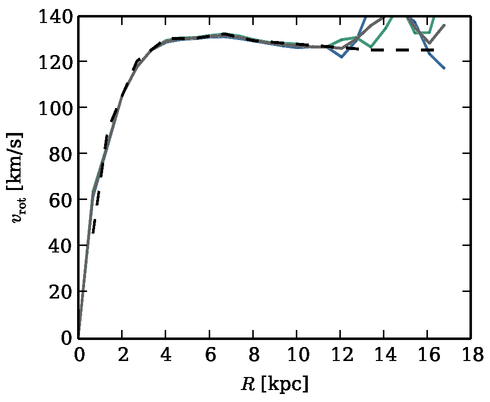}
\includegraphics[width=0.32\textwidth]{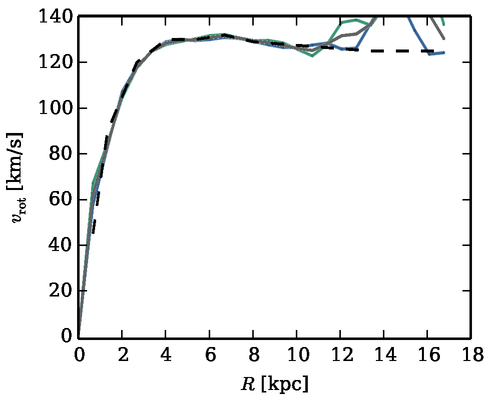}
\includegraphics[width=0.32\textwidth]{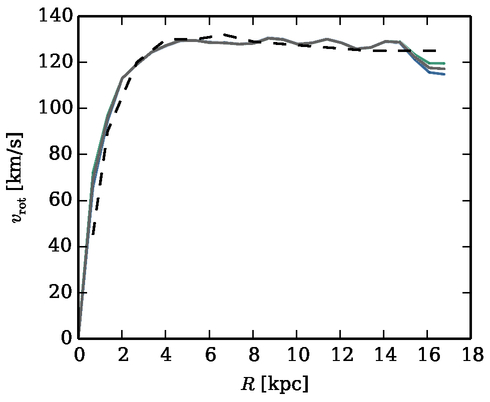}
\includegraphics[width=0.32\textwidth]{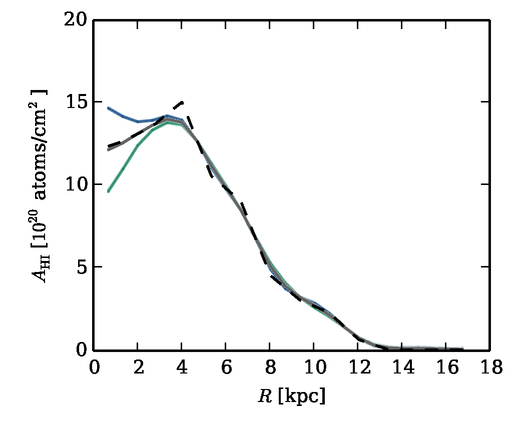}
\includegraphics[width=0.32\textwidth]{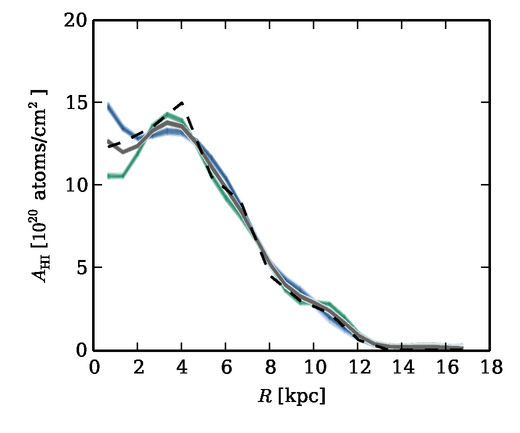}
\includegraphics[width=0.32\textwidth]{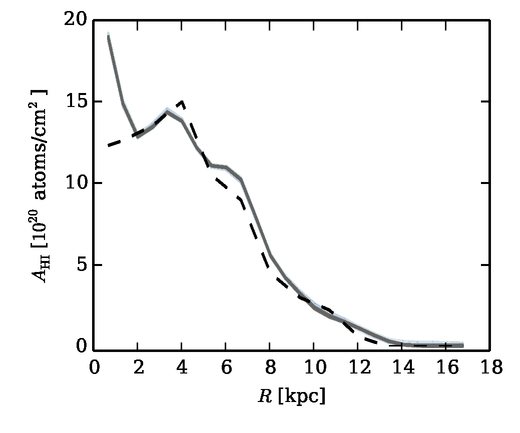}
\includegraphics[width=0.32\textwidth]{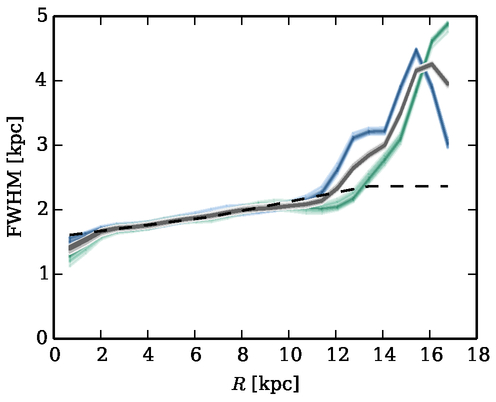}
\includegraphics[width=0.32\textwidth]{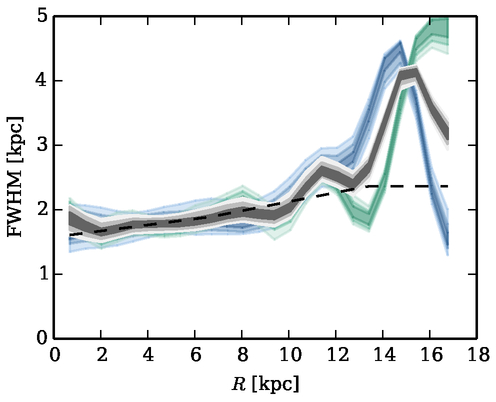}
\includegraphics[width=0.32\textwidth]{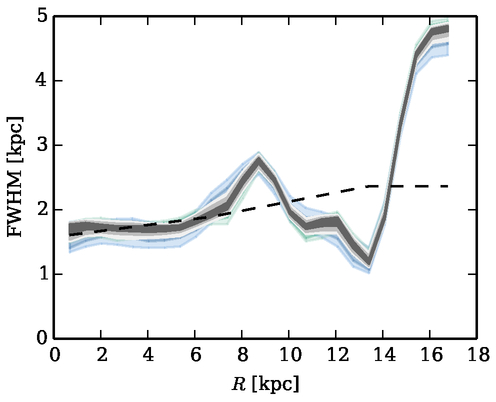}
\includegraphics[width=0.32\textwidth]{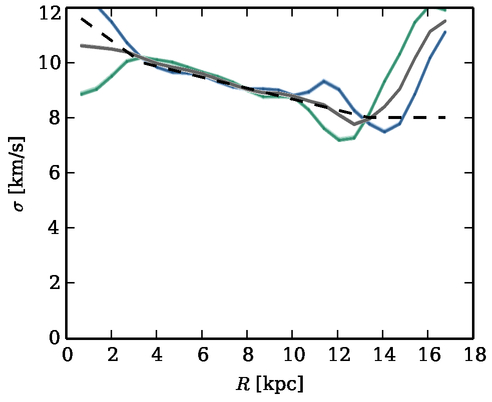}
\includegraphics[width=0.32\textwidth]{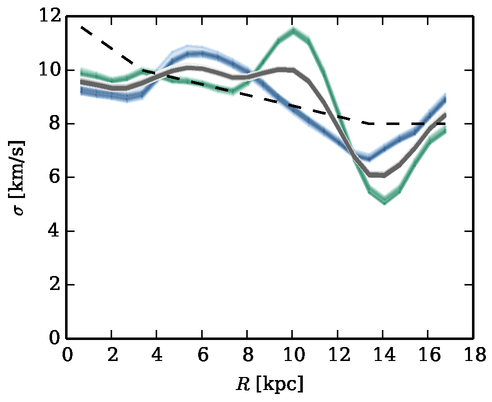}
\includegraphics[width=0.32\textwidth]{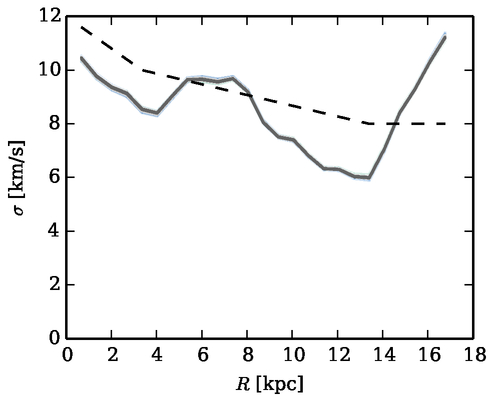}
\caption[Self-absorption test model fits (1\,K, 3\,K and 5\,K)]{Self-absorption ($T_\textrm{spin}=100$\,K) test models with noises of 1\,K, 3\,K and 5\,K (left-to-right). Top panels shows the rotation curve $v_\textrm{rot}$, second row shows the face-on surface density $A_\hi$. Third row shows the thickness of the model. Fourth row shows the fit to the velocity dispersion.  Green colours denote the left side of the galaxy, blue colours show the right side. The grey band is the combined result. Dashed lines show the true parameters of the model.}\label{fig:testmodels-thick}
\end{figure*}

\subsection{Self-absorption Fits}
We perform a similar test using these models, but include 
self-absorption at a median spin temperature of 100\,K.
The face-on surface density is increased to compensate for 
the self-absorption and now peaks around $1.5\times10^{20}$\,atoms/cm$^2$.
We show the results in Figure \ref{fig:testmodels-thick}. 
We find that the rotation curve is recovered well in all 
three models.
Both the measurements of face-on surface density and those of
the velocity dispersion become
increasingly difficult at lower signal-to-noise ratio.
The velocity dispersion is only recovered well in the 1\,K model.

\begin{figure*}
\centering
\includegraphics[width=0.32\textwidth]{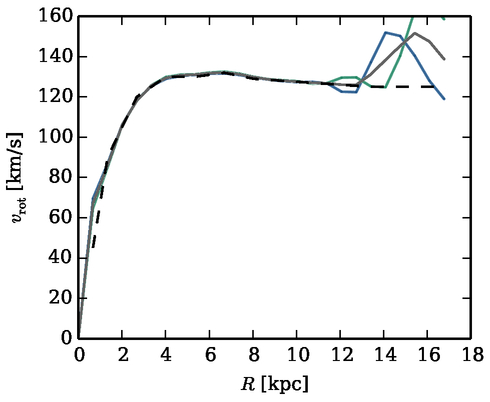}
\includegraphics[width=0.32\textwidth]{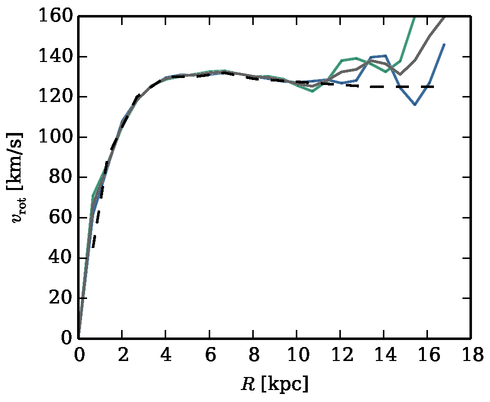}
\includegraphics[width=0.32\textwidth]{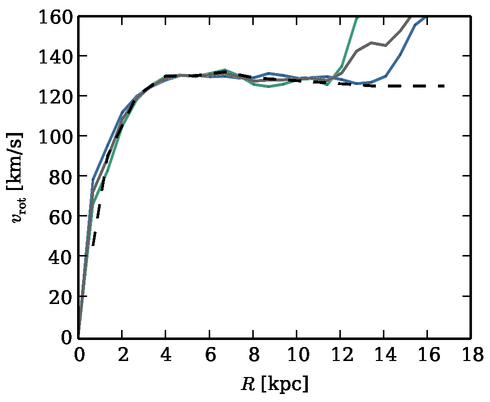}
\includegraphics[width=0.32\textwidth]{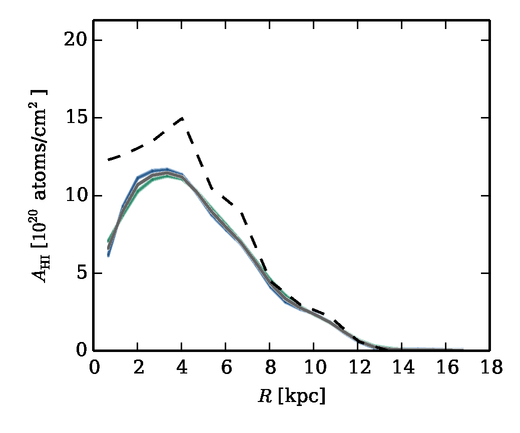}
\includegraphics[width=0.32\textwidth]{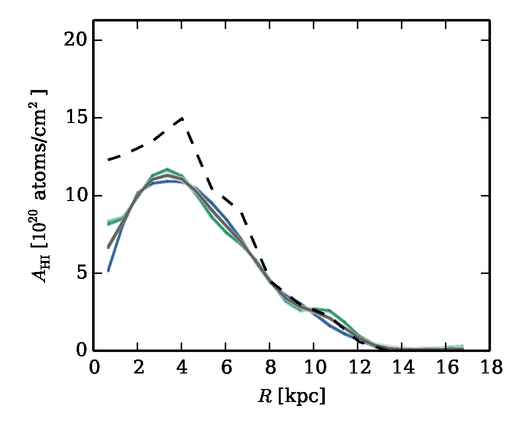}
\includegraphics[width=0.32\textwidth]{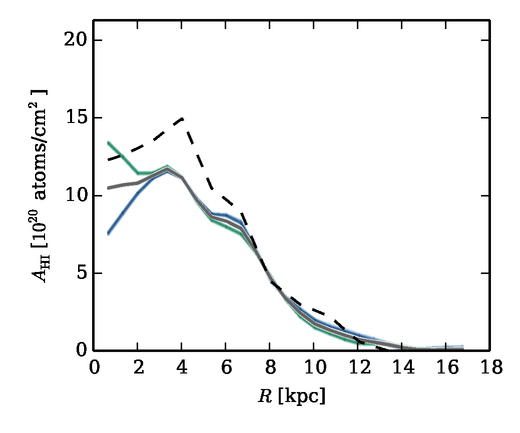}
\includegraphics[width=0.32\textwidth]{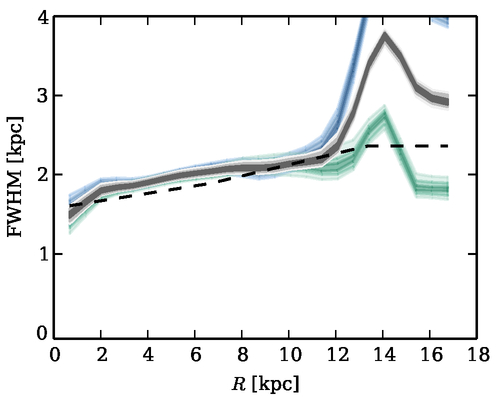}
\includegraphics[width=0.32\textwidth]{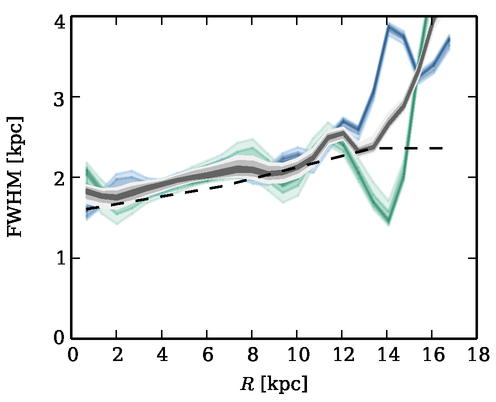}
\includegraphics[width=0.32\textwidth]{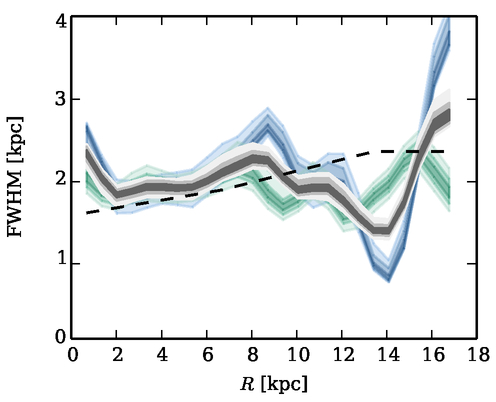}
\includegraphics[width=0.32\textwidth]{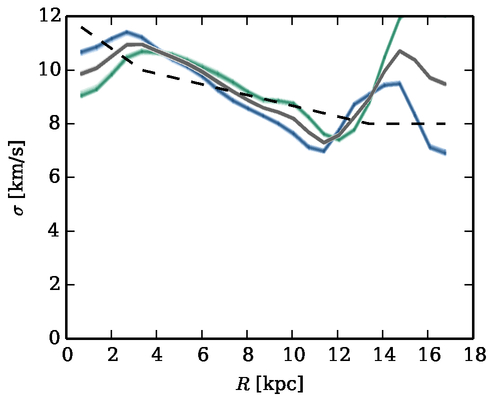}
\includegraphics[width=0.32\textwidth]{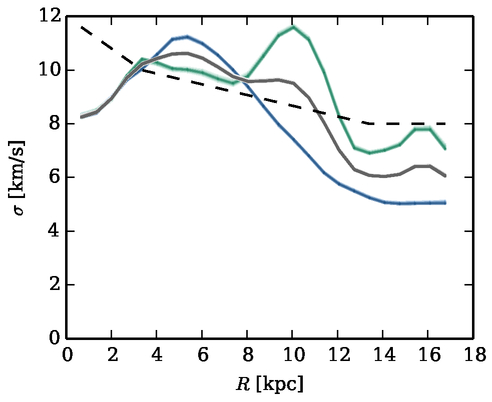}
\includegraphics[width=0.32\textwidth]{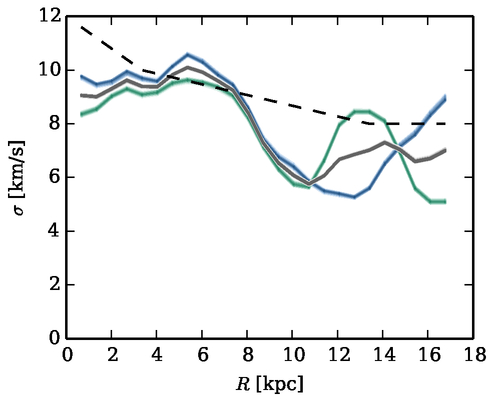}
\caption[Optically thin fits to self-absorption models]{Optically thin fits to self-absorption ($T_\textrm{spin}=100$\,K) test models with noises of 1\,K, 3\,K and 5\,K (left-to-right). Top panels shows the rotation curve $v_\textrm{rot}$, second row shows the face-on surface density $A_\hi$. Third row shows the thickness of the model. Fourth row shows the fit to the velocity dispersion.  Green colours denote the left side of the galaxy, blue colours show the right side. The grey band is the combined result. Dashed lines show the true parameters of the model.}\label{fig:HIwhenmodelsgobad}
\end{figure*}

\begin{figure*}
\centering
\includegraphics[width=0.32\textwidth]{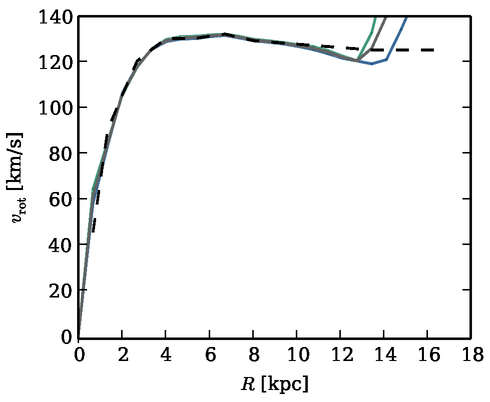}
\includegraphics[width=0.32\textwidth]{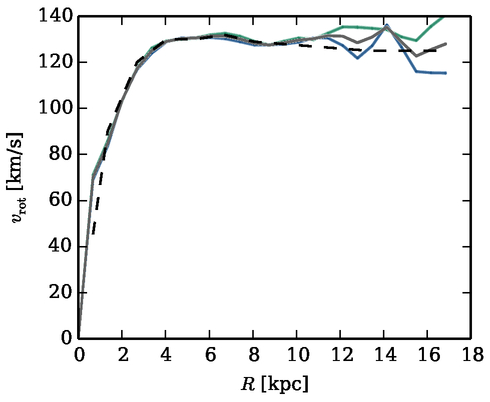}
\includegraphics[width=0.32\textwidth]{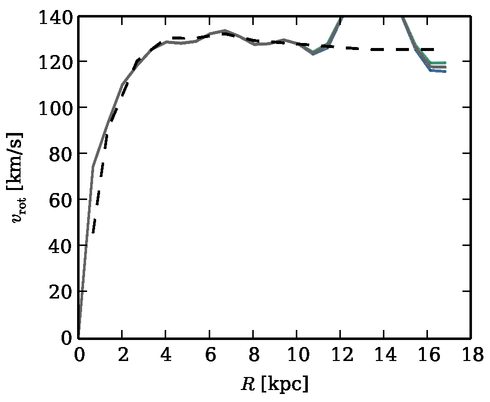}
\includegraphics[width=0.32\textwidth]{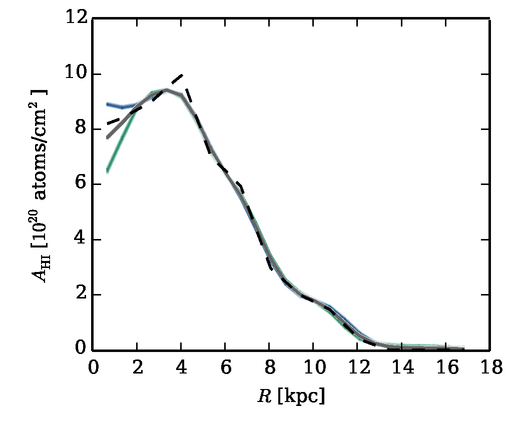}
\includegraphics[width=0.32\textwidth]{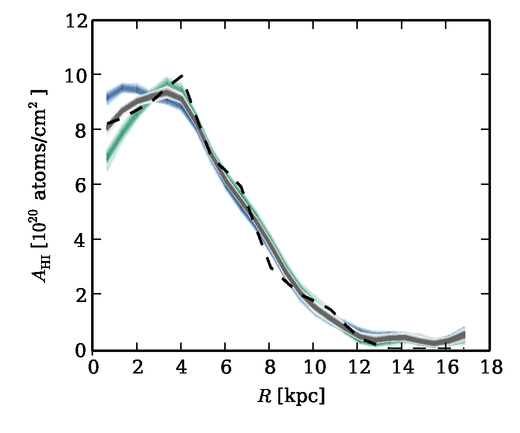}
\includegraphics[width=0.32\textwidth]{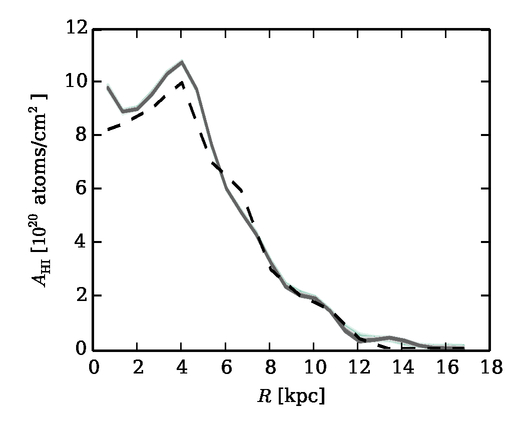}
\includegraphics[width=0.32\textwidth]{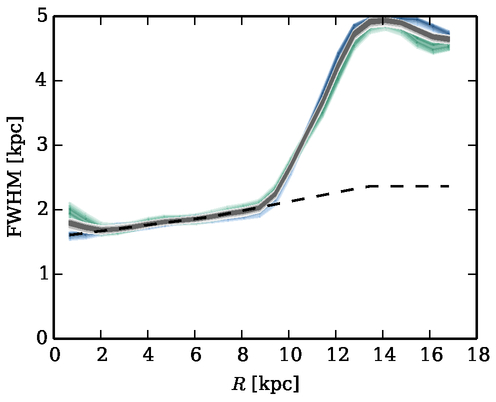}
\includegraphics[width=0.32\textwidth]{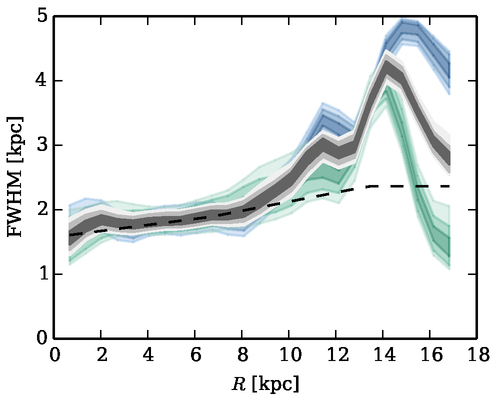}
\includegraphics[width=0.32\textwidth]{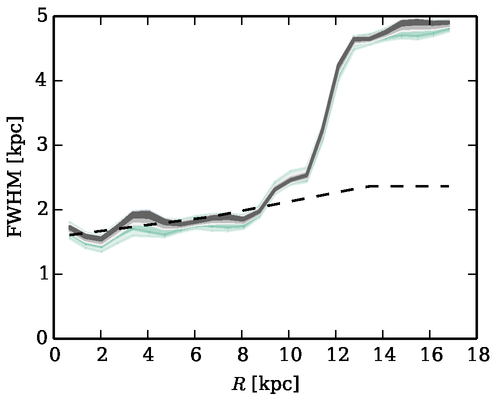}
\includegraphics[width=0.32\textwidth]{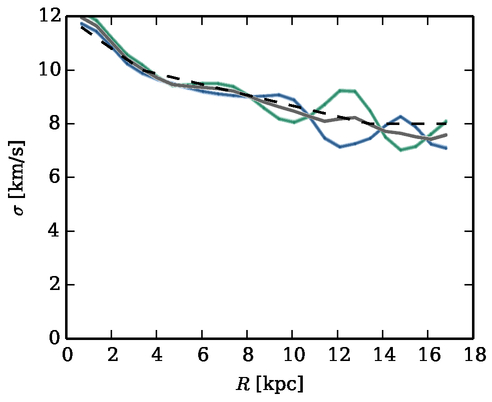}
\includegraphics[width=0.32\textwidth]{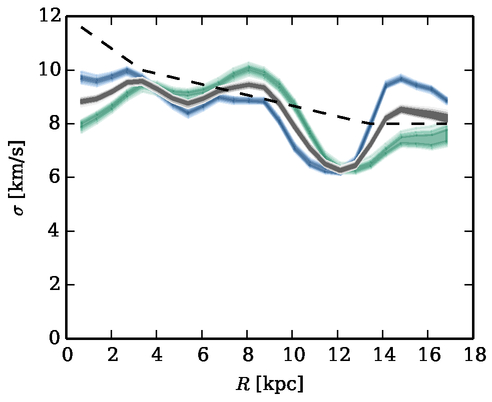}
\includegraphics[width=0.32\textwidth]{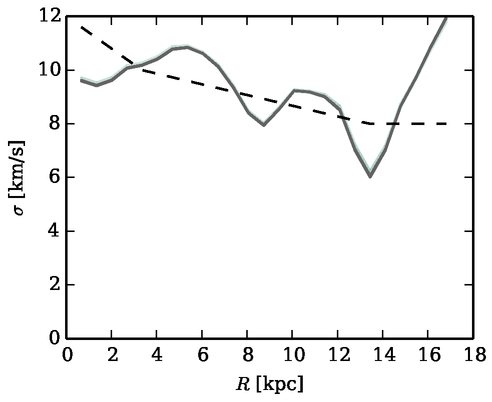}
\caption[Optically thin test model fits at $i=88.8^\circ$ (1\,K, 3\,K and 5\,K)]{Optically thin test models with noises of 1\,K, 3\,K and 5\,K (left-to-right). The reference images have an inclination of $88.8^\circ$. Top panels shows the rotation curve $v_\textrm{rot}$, second row shows the face-on surface density $A_\hi$. Third row shows the thickness of the model. Fourth row shows the fit to the velocity dispersion.  Green colours denote the left side of the galaxy, blue colours show the right side. The grey band is the combined result. Dashed lines show the true parameters of the model.}\label{fig:testmodels-wronginclination}
\end{figure*}

\subsection{A Self-absorption Model Fit as Optically Thin}
If in reality the \hi is self-absorbing, how wrong will models be
that assume an 
optically-thin medium?
To test this, we fit the self-absorption models from the previous test with 
optically-thin models.
The results are shown in Figure 
\ref{fig:HIwhenmodelsgobad}.\label{sec:whenmodelgobad}
Again the rotation curve is recovered well out to 12 kpc or so. 
Clearly, the rotation curve does not suffer much from self-absorption. 
As expected, the face-on surface density is not 
recovered well at all, with the highest column densities estimated at 
only $\sim70$\% of their true values.
The disc is consistently estimated to be somewhat
thicker than it in reality is.
The velocity dispersion is not recovered well at all.
Performing optically thin fits on self-absorbed distributions produces
results significantly in error, even at relatively small radii.

\subsection{A More Complicated Model}\label{sec:notedgeon}
In the previous tests we have modelled and fitted the galaxies at 
a perfect edge-on inclination of $90^\circ$. \label{sec:whenHIgoesbad}
Our fitting strategy always assumes this perfectly edge-on inclination.
As a final test, we examine the quality of the fit if the actual 
galaxy is not perfectly edge-on, but is at an inclination of $88.8^\circ$.
We also include a modest side-on warp that begins at $R=8$\,kpc.
We again run the fits with intrinsic noises of 1, 3 and 5\,K 
and use optically thin models.
The results are shown in Figure \ref{fig:testmodels-wronginclination}. 
The true systemic velocity is recovered to within 0.6\,km/s, 
while the channel width was 3.3\,km/s.
The position angle is accurate to within $0.9^\circ$.
The central position is found to within 0.7''.
The measurement of the thickness of the disc is wrong beyond 
a radius of 9\,kpc, which is due to the onset of the warp.
Our conclusions remain the same as the previous tests: All 
parameters are recovered well at 1\,K, but the measurements of the 
velocity dispersion become increasingly difficult at 3\,K and for
all parameters at 5\,K.

\subsection{Concluding Remarks}
So what can we conclude? 
We have shown that our fitting strategy can recover the parameters 
in these edge-on galaxies.
The rotation curve and the face-on surface density are recovered 
well in all cases.
The biggest problem in fitting these galaxies comes from the velocity 
dispersion, which can only be fit in very low noise  \hi data cubes.
We will thus adopt a constant velocity dispersion for galaxies where 
the noise is too high.
The flaring will only be problematic in high noise  \hi data cubes.
We note that our strategy produces these results in idealized circumstances.
In reality, the galaxies can (and will) have asymmetries, spiral arms, 
warps, lagging haloes, etc.; which will make our measurements less accurate 
then the presented test models.

We have also shown that fitting a self-absorption reference model with an 
optically thin model will lead to bad results. 
In particular, the face-on surface density will be too low, while the 
thickness measured will be too high and the velocity dispersion will be wrong.
Only the rotation curve will be estimated to a reasonable degree.
As we argued before, galaxies are in reality expected to have self-absorbing 
\hi. 
This result thus implies that the results found in the 
literature based on the assumption
of zero optical thickness are underestimating the true \hi content of
galaxies.

\section{Fits to actual galaxies}\label{sec:HIfittingresults}
In Paper I, we have presented the \hi observations 
for eight edge-on galaxies. 
In this section, we will focus on modelling these galaxies in more detail.
We aim to fit the rotation curve, velocity dispersion, thickness 
and face-on surface density of the galaxies, as we will use these 
in Paper V to model the hydrostatics of the gas.
Based on the results of the previous section, it is of vital 
importance to have the lowest noise levels we possibly can.
To this end, we re-image the data such that we can still resolve 
the vertical structure of the disc, but have the lowest noise possible.
For more details on the \hi reduction, see Section 4.1 of Paper I.
In Table 3 of Paper I, we denote the original  \hi data cubes.
IC\,5052 has been re-imaged with a FWHM beam of 25'' and a noise 
$\sigma$ of 1.7\,K.
ESO\,115-G021 now has a FWHM beam of 25'' and a noise $\sigma$ of 1.27\,K.
For ESO\,274-G001, we adopt the 30'' beam from Table 
3 of Paper I, which has a noise $\sigma$ of 1.2\,K.
UGC\,7321 has been re-imaged with a FWHM beam of 25'' and a 
noise $\sigma$ of 1.7\,K.
Based on these noise levels and the results of the previous section, 
we have decided to model UGC\,7321, ESO\,115-G021 and ESO\,274-G001 
with the velocity dispersion as a free parameter that can vary with radius.
The other galaxies will be modelled using an assumed constant velocity 
dispersion of 10\,km/s.

We do not include the results for IC\,2531 and ESO\,146-G014. 
In Section 5.1 of Paper I, we commented on the remarkable thin 
disc that IC\,2531 appeared to have.
This disc, combined with the high noise, made it impossible to  
measure the thickness of the \hi disc accurately.
We were unable to make a reliable fit to ESO\,146-G014.
As we discussed in Section 5.6 of Paper I, the galaxy probably 
has a warp and is not sufficiently edge-on for our measurements.

\subsection{IC\,5052}
\begin{figure*}
\centering
   \includegraphics[width=0.32\textwidth]{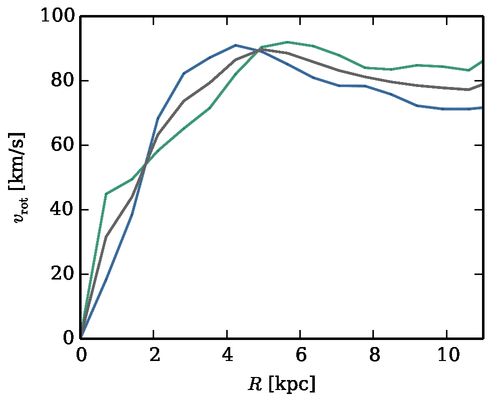}
   \includegraphics[width=0.32\textwidth]{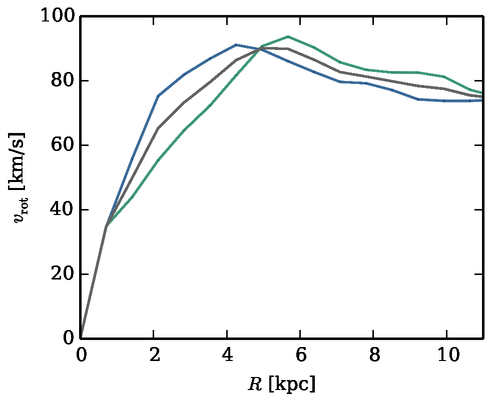}
   \includegraphics[width=0.32\textwidth]{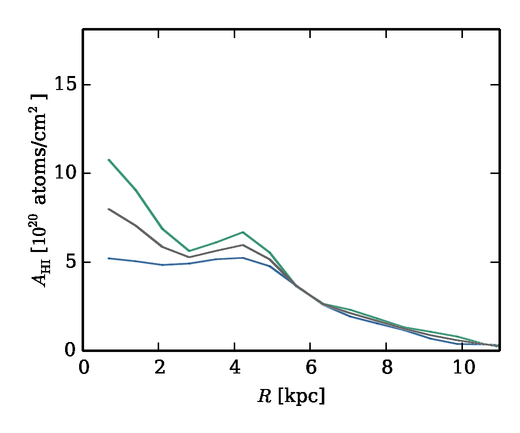}
   \includegraphics[width=0.32\textwidth]{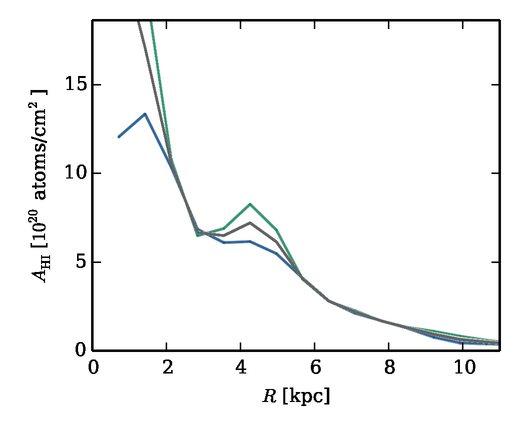}
   \includegraphics[width=0.32\textwidth]{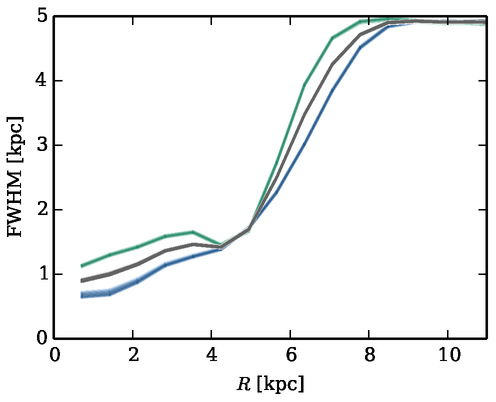}
   \includegraphics[width=0.32\textwidth]{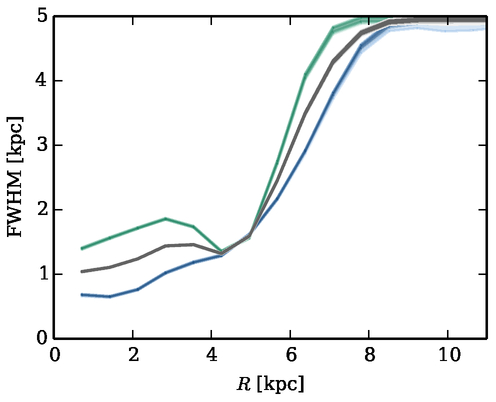}
\caption[\hi decomposition of IC\,5052]{Decomposition results for IC\,5052 assuming an optically thin mode (left column) and self-absorbing \hi medium (right column). Colour ranges have been chosen such that they contain 68\%, 95\% and 99.7\% of the distribution. Assuming a normal distribution, this represents the 1, 2 and 3$\sigma$ dispersion from the mean. Green lines are for the left side of the galaxy, cyan lines for the right side.}\label{fig:IC5052-velocitydispersion-thin}
\end{figure*}

Galaxy IC\,5052 has proven hard to model.
We show both the optically thin model and the self-absorption model in 
Figure \ref{fig:IC5052-velocitydispersion-thin}.
As we noted in Section 5.2 of Paper I, the galaxy has a very strong warp. 
Looking back at Figure 3 in Paper I, we can see that 
the onset of the warp occurs just beyond that same radius (180'').
This warp is responsible for the very large thickness that is found 
beyond 5\,kpc.
The declining rotation curve beyond 5\,kpc is also due to this component.

The self-absorption models have a total mass of 
$9.5\pm0.9\times10^8$\,M$_\odot$, while the optically thin model recovers 
only $7.4\pm0.7\times10^8$\,M$_\odot$. 
This is less than our initial estimate of $8.9\times10^8$\,M$_\odot$ 
in Table 4 of Paper I.
This trend is also visible in the other galaxies. It is because the 
fit can only model a smooth medium. Local bright regions in the 
observations therefore cannot be accurately recovered.
Any warp will also harbor mass, which is not recovered here.
Comparing the two models (and ignoring the right-hand self-absorption 
model), the optically thin model is missing about a quarter of the \hi.

\citet{OBrien2010C} also had trouble with modelling IC\,5052.
They did not resolve the warp, but found clear evidence for asymmetry.
Comparing the rotation curves to our work, the results on both sides 
look similar up to 6\,kpc, but subsequent downturn is not detected as 
strongly in their work. The thickness estimates are similar. 
Their face-on column density is beyond 2\,kpc similar to our optically 
thin result, although due to our smoothing the data is less bumpy.
The inner parts are different, but are not well resolved in our models. 
The results we find for the flaring beyond a radius of 5 kpc are doubtful. 
We have chosen in Paper V in this series where we analyze our data further 
not to consider this galaxy.

%&*&

\begin{figure*}
\centering
   \includegraphics[width=0.32\textwidth]{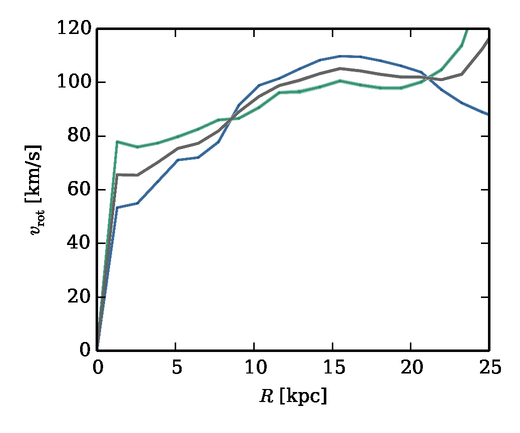}
   \includegraphics[width=0.32\textwidth]{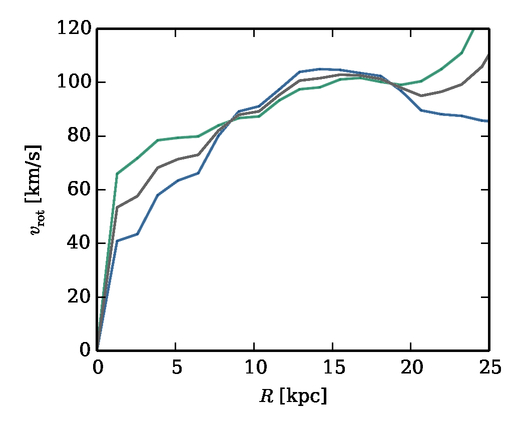}
   \includegraphics[width=0.32\textwidth]{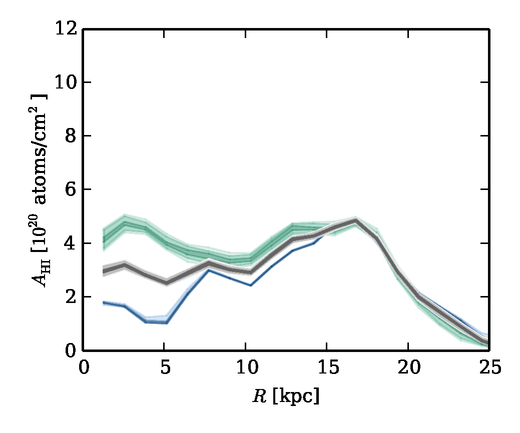}
   \includegraphics[width=0.32\textwidth]{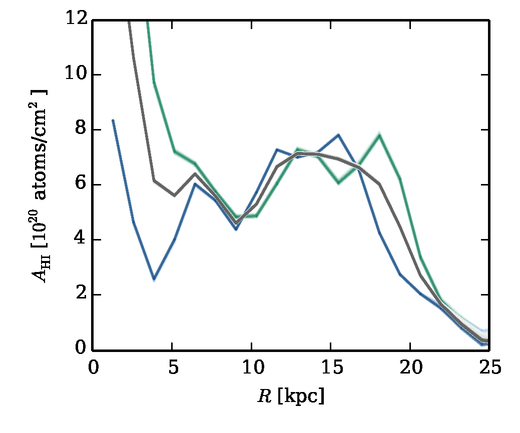}
   \includegraphics[width=0.32\textwidth]{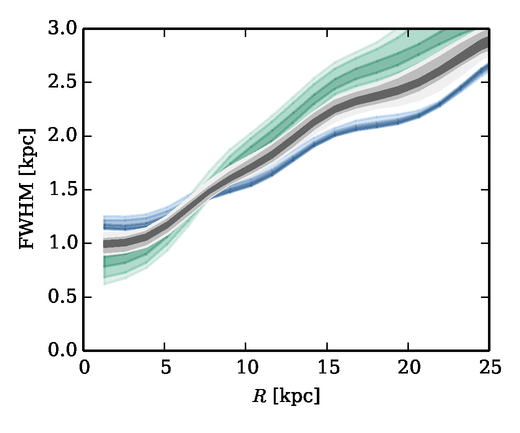}
   \includegraphics[width=0.32\textwidth]{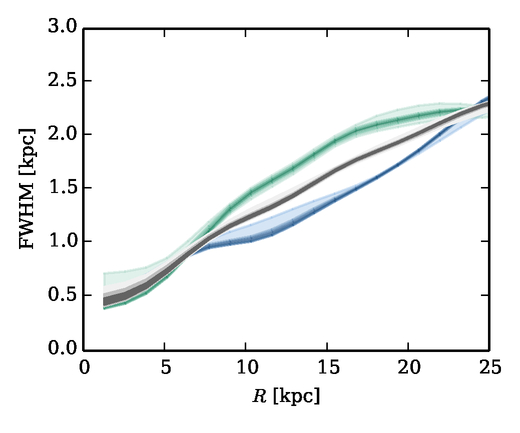}
\caption[\hi decomposition of IC\,5249 (optically thin)]{Decomposition results for IC\,5249 assuming an optically thin mode (left column) and self-absorbing \hi medium (right column). Colour ranges have been chosen such that they contain 68\%, 95\% and 99.7\% of the distribution. Assuming a normal distribution, this represents the 1, 2 and 3$\sigma$ dispersion from the mean. Green lines are for the left side of the galaxy, cyan lines for the right side.  The grey band is the combined result.}  \label{fig:IC5249-velocitydispersion-thin}
\end{figure*}

\begin{figure*}
   \includegraphics[width=0.32\textwidth]{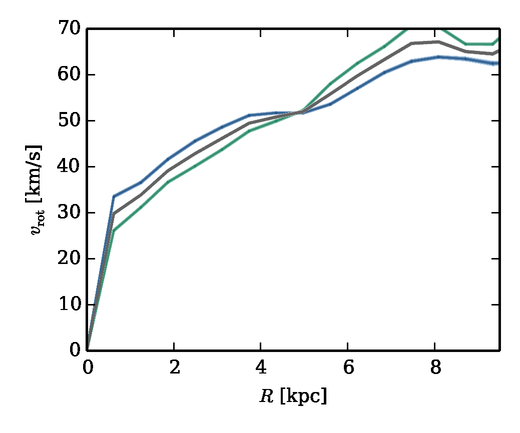}
   \includegraphics[width=0.32\textwidth]{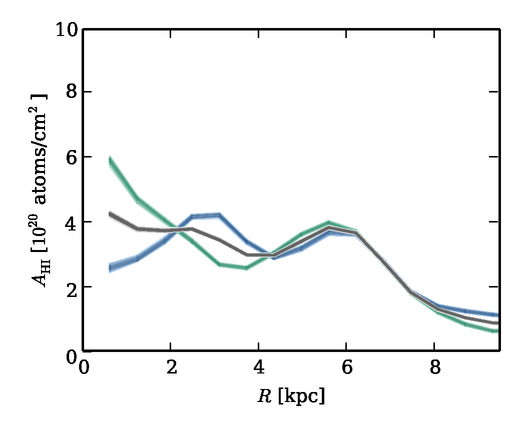}

   \includegraphics[width=0.32\textwidth]{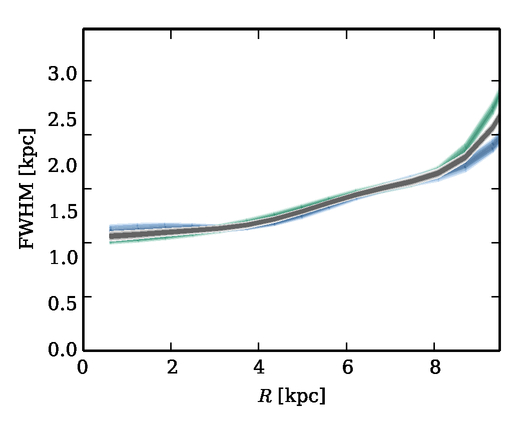}
   \includegraphics[width=0.32\textwidth]{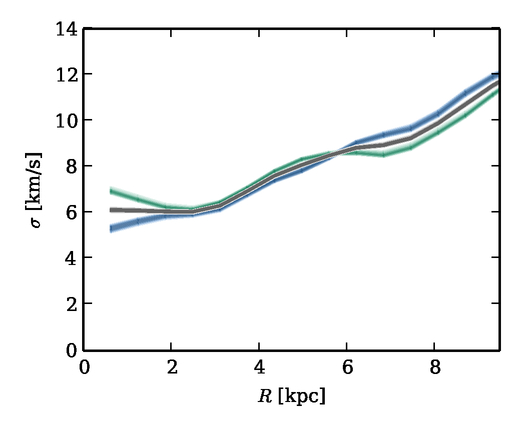}
\caption[\hi decomposition of ESO\,115-G001 (optically thin)]{Decomposition results for ESO\,115-G021 assuming an optically thin model. Colour ranges have been chosen such that they contain 68\%, 95\% and 99.7\% of the distribution. Assuming a normal distribution, this represents the 1, 2 and 3$\sigma$ dispersion from the mean. Green lines are for the left side of the galaxy, cyan lines for the right side.  The grey band is the combined result.}
   \label{fig:ESO115-G021-velocitydispersion-thin}
\end{figure*}

\begin{figure*}
   \includegraphics[width=0.32\textwidth]{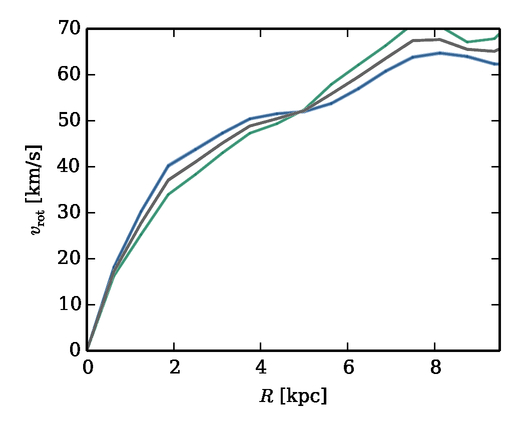}
   \includegraphics[width=0.32\textwidth]{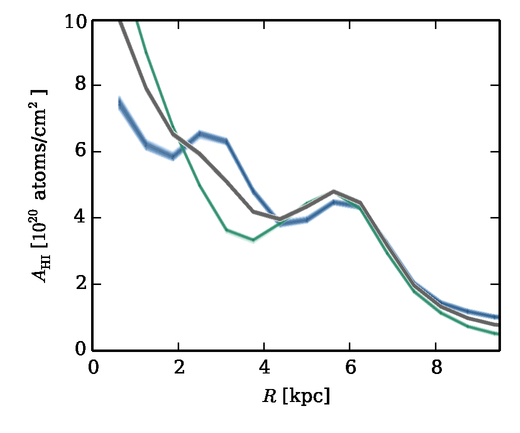}

   \includegraphics[width=0.32\textwidth]{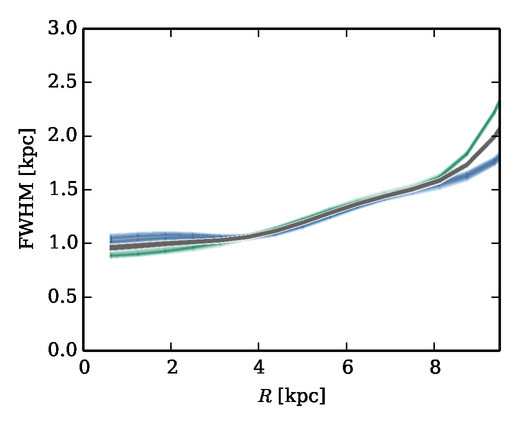}
   \includegraphics[width=0.32\textwidth]{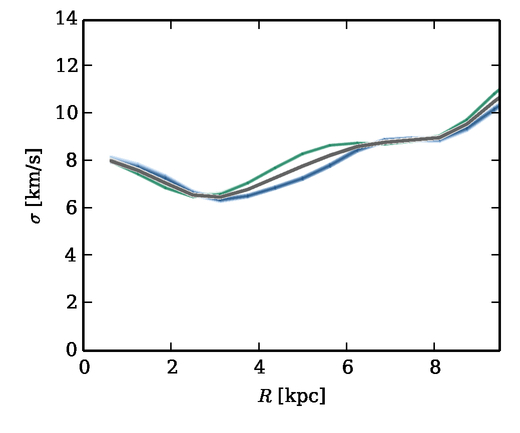}
\caption[\hi decomposition of ESO\,115-G021 (self-absorbing)]{Decomposition results for ESO\,115-G021 assuming a self-absorption model. Colour ranges have been chosen such that they contain 68\%, 95\% and 99.7\% of the distribution. Assuming a normal distribution, this represents the 1, 2 and 3$\sigma$ dispersion from the mean. Green lines are for the left side of the galaxy, cyan lines for the right side.  The grey band is the combined result.}
   \label{fig:ESO115-G021-velocitydispersion-thick}
\end{figure*}

\begin{figure*}
\centering
   \includegraphics[width=0.32\textwidth]{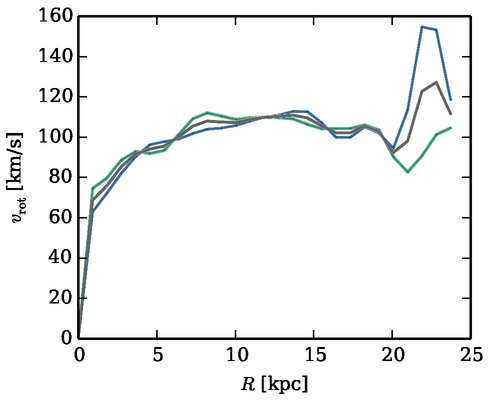}
   \includegraphics[width=0.32\textwidth]{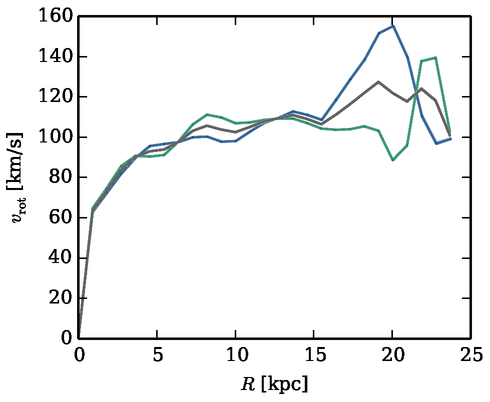}
   \includegraphics[width=0.32\textwidth]{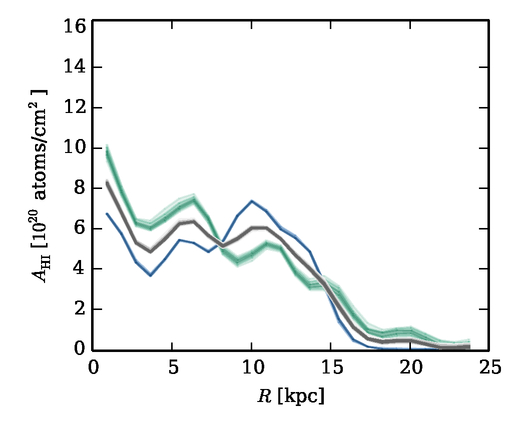}
   \includegraphics[width=0.32\textwidth]{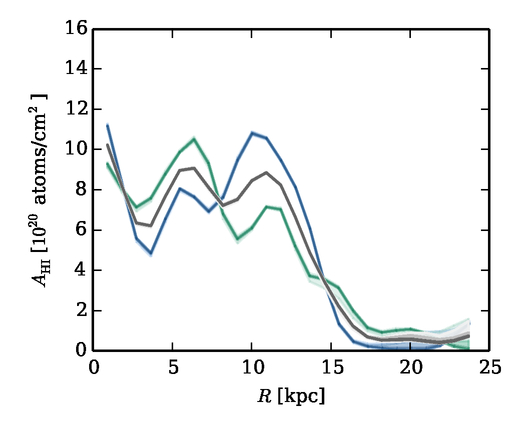}
   \includegraphics[width=0.32\textwidth]{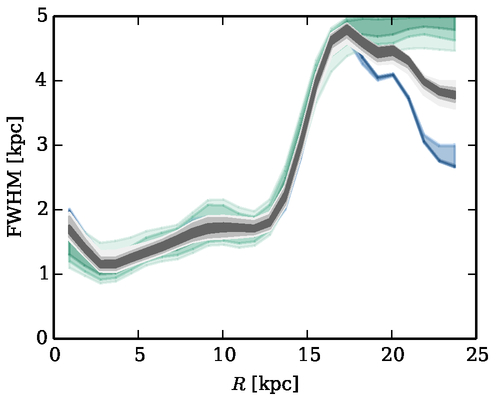}
   \includegraphics[width=0.32\textwidth]{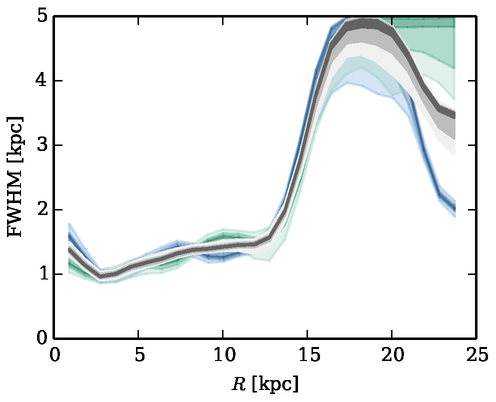}
\caption[\hi decomposition of ESO\,138-G014 (optically thin)]{Decomposition results for ESO\,138-G014 assuming  an optically thin mode (left column) and self-absorbing \hi medium (right column). Colour ranges have been chosen such that they contain 68\%, 95\% and 99.7\% of the distribution. Assuming a normal distribution, this represents the 1, 2 and 3$\sigma$ dispersion from the mean. Green lines are for the left side of the galaxy, cyan lines for the right side.  The grey band is the combined result.}\label{fig:ESO138-G014-velocitydispersion-thin}
\end{figure*}

\subsection{IC\,5249}
The \hi structure and kinematics of galaxy IC\,5249 have been previously 
analyzed by \citet{Abe1999},  \citet{vanderKruit2001A} and \citet{OBrien2010C}.
\citet{Abe1999} report a linearly rising rotation curve to about 100\,km/s 
at 17\,kpc.
The data was re-analyzed by \citet{vanderKruit2001A}, who reports a steeper 
inner rotation curve that flattens out to 105 km/s at 10\,kpc and remains 
flat throughout the remaining disc.
This rotation curve was confirmed by \citet{OBrien2010C}.
Our fits with both the optically thin model and the self-absorption model  
confirms their results  (Figure \ref{fig:IC5249-velocitydispersion-thin}).
We however note that rotation curves differs between both sides, so the 
galaxy might be asymmetric.

Comparing the face-on surface densities, our optically thin fit looks very 
similar to the observed face-on density by \citet{vanderKruit2001A}. The 
peak surface density occurs at a radius of 17-18\,kpc at a value of 
$5.8\times10^{20}$ atoms/cm$^2$. More inward, the profile has a lower 
density. We also reproduce their dip in the face-on profile at $\sim11$\,kpc.
The face-on surface density of \citet{OBrien2010C} looks similar to 
our left-side profile.
Their joined profile does recover the peak at 17-18\,kpc, but does 
not have the lower central densities.
Our self-absorption fit finds more \hi. The peak at 17-18\,kpc 
now becomes a plateau near at $8\times10^{20}$\,atoms/cm$^2$ (30\% higher). 
The profile also shows the dip near 11\,kpc, although the 
average inner density remains far larger than the optically thin result. 
The inner parts of the optically thin model are `saturated' 
by the \hi from outer radii, and the model compensates for 
this by lowering the face-on column densities in the inner parts.

We find a strongly flaring disc in both fits, with an inner 
thickness of only 500\,pc, but increasing linearly with radius 
to nearly 2.6\,kpc at 25\,kpc. In Paper V we will only use the flaring 
out to 14 kpc for the fits to determine the axis ratio of the dark matter 
halo.
This is different than \citet{OBrien2010C}, who finds the same 
linearly increasing thickness, but only going out to 1.5\,kpc 
thickness at a radius of 20\,kpc.
An average thickness of $1.1\pm0.3$\,kpc was reported by 
\citet{vanderKruit2001A}, although their Figure 6 can be 
seen to increase linearly to a thickness of 3.3\,kpc at 14\,kpc. 
Given their error bars, this result is consistent with our findings.

Comparing the total masses of both types of fits, we find 
that the optically thin model has a total \hi mass of 
$4.8\pm0.2\times10^9$\,M$_\odot$.
The self-absorption model finds a total of $7.8\pm0.8\times10^9$\,M$_\odot$.
This would put the hidden \hi mass fraction at about a third.
The optically thin mass is however significantly lower than our 
initial estimate of $5.6 \times 10^{9}$\,M$_\odot$ 
(Table 4 of Paper I). 

\subsection{ESO\,115-G021}
Galaxy ESO\,115-G021 is a symmetric, slow rotating 
galaxy which we have managed to fit well (see Figures 
\ref{fig:ESO115-G021-velocitydispersion-thin} 
and \ref{fig:ESO115-G021-velocitydispersion-thick}).
The rotation curve rises from an initial 30\,km/s 
in the inner parts almost linearly to 65\,km/s around 8\,km/s.
The only other analysis of the rotation curve of 
this galaxy is by \citet{OBrien2010C}, who report a very similar rotation curve.
The face-on surface density of our optically thin 
model is also similar to theirs: A plateau between 
$3-5\times10^{20}$\,atoms/cm$^2$ and then rapidly dropping off.
The self-absorption fit to the surface density yields 
a similar profile, although the inner part has a bit more \hi.

The thickness of this galaxy is remarkably well behaved 
and consistent between both fits.
Starting  near 700\,pc in the inner parts, it rises 
linearly to 1.5\,kpc at a radius of 8\,kpc.
This is different to \citet{OBrien2010C}, who report 
a stronger increase  to over 3\,kpc beyond a radius 
of 6\,kpc, although they note they had insufficient available 
positions in their data for a proper fit.

The velocity dispersion fit to the optically thin model is 
surprising; the galaxy appears to have an increasing velocity 
dispersion, which is 7\,km/s in the inner parts and 10\,km/s at 6\,kpc.
The self-absorption velocity dispersion is more stable, although 
it is still varying a bit. 
This behavior is likely due to noise in the data.

Comparing the total masses, the optically thin model yields 
$5.6\pm0.2\times10^9$\,M$_\odot$ and $7.2\pm0.1\times10^9$\,M$_\odot$ 
for the self-absorption model.
Thus about a quarter of the \hi is hidden in the model.

\subsection{ESO\,138-G014}\label{sec:HIdecom138}
Galaxy ESO\,138-G014 has been well resolved.
The optically thin model and  the self-absorption 
model are shown in Figure \ref{fig:ESO138-G014-velocitydispersion-thin}.
The rotation curve rises steeply to 70\,km/s in the 
inner parts, and gradually increase to a maximum of 105\,km/s near 8\,kpc. 
Beyond this radius, it levels off.
The curve is very similar to that measured by \citet{OBrien2010C}.

The disc is reasonably thick, starting  at 1.1\,kpc 
near 2.5\,kpc radius and increasing linearly towards 
1.8\,kpc at 9\,kpc (for the optically thin model).
In the self-absorption model, the disc is less thick, 
starting around 1\,kpc in the inner parts and increasing 
only to 1.5\,kpc at 9\,kpc.
In both models, around 12-13\,kpc the thickness suddenly 
increases drastically, almost certainly due to the warp
seen in the total \hi map (Fig. 7 in Paper I).
Note that the increase towards a thickness of 2\,kpc in 
the centre is unreliable.
\citet{OBrien2010C} do not trust their thickness 
measurements and so we cannot compare their results with ours.
In Paper V we will only use the flaring 
out to 6 kpc for the fits to determine the axis ratio of the dark matter 
halo.

In the optically thin model, the face-on surface 
density shows a clear plateau between 5 and $6\times10^{20}$\,atoms/cm$^2$ 
up to a radius of 12-13\,kpc, and dropping off rapidly beyond that.
The self-absorption model results in more \hi, with the plateau at a 
higher $6-9\times10^{20}$\,atoms/cm$^2$.
The (optically thin) model of \citet{OBrien2010C} also finds a plateau 
near $6\times10^{20}$\,atoms/cm$^2$, although it is higher near the inner parts.

Comparing the two mass estimates, our optically thin model has a total 
mass of $3.4\pm0.1 \times 10^9$\,M$_\odot$, while our self-absorption model 
finds $4.6\pm 0.2 \times 10^9$\,M$_\odot$.
The total hidden \hi fraction  is thus $28\pm1$\% of the total \hi.

\begin{figure*}
   \includegraphics[width=0.32\textwidth]{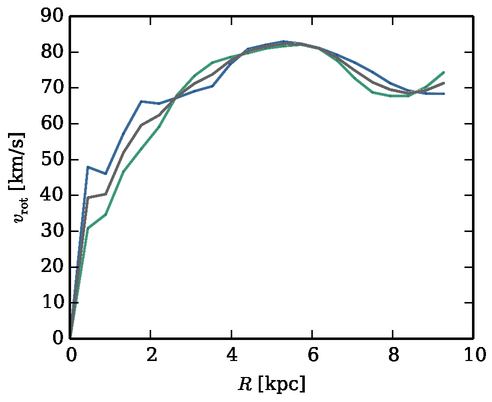}
   \includegraphics[width=0.32\textwidth]{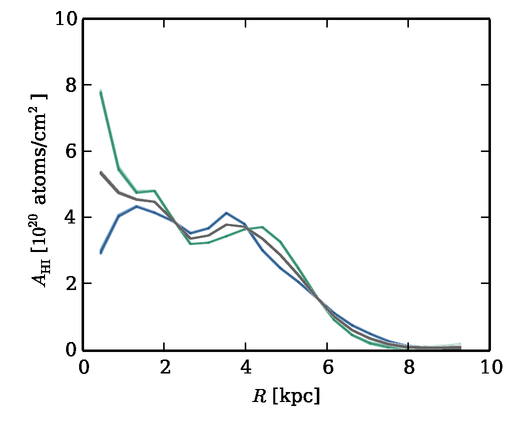}

   \includegraphics[width=0.32\textwidth]{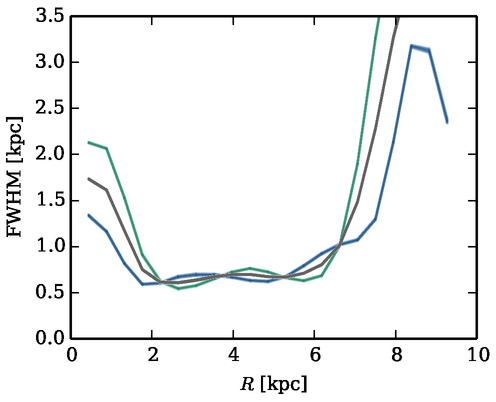}
   \includegraphics[width=0.32\textwidth]{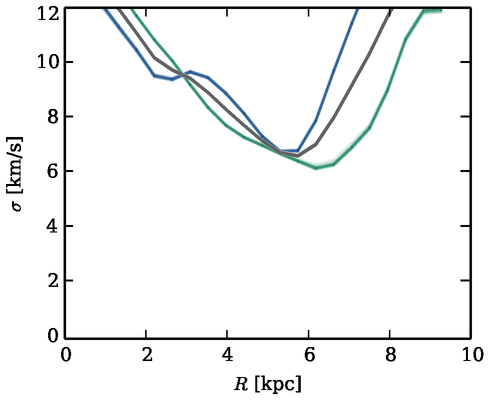}
\caption[\hi decomposition of ESO\,274-G001 (optically thin)]{Decomposition results for ESO\,274-G001 assuming an optically thin model. Colour ranges have been chosen such that they contain 68\%, 95\% and 99.7\% of the distribution. Assuming a normal distribution, this represents the 1, 2 and 3$\sigma$ dispersion from the mean. Green lines are for the left side of the galaxy, cyan lines for the right side.  The grey band is the combined result.}
   \label{fig:ESO274-G001-velocitydispersion-thin}
\end{figure*}

\begin{figure*}
   \includegraphics[width=0.32\textwidth]{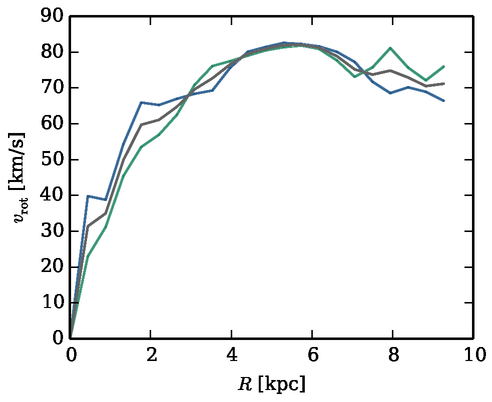}
   \includegraphics[width=0.32\textwidth]{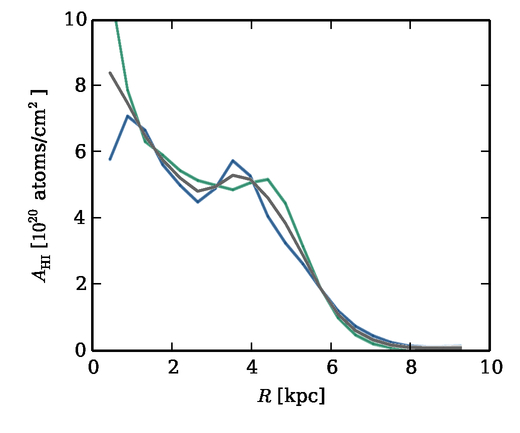}

   \includegraphics[width=0.32\textwidth]{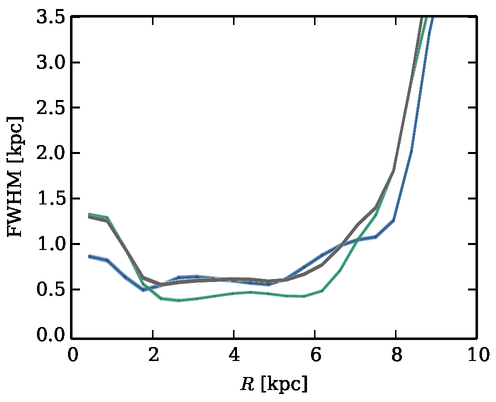}
   \includegraphics[width=0.32\textwidth]{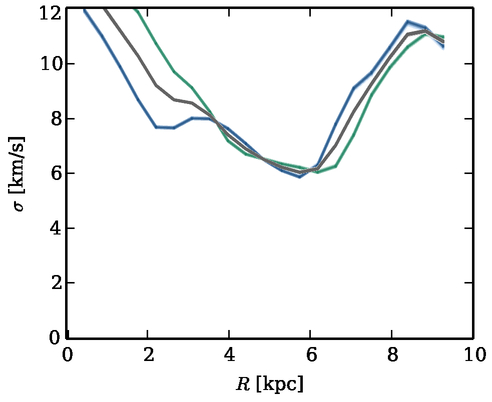}
\caption[\hi decomposition of ESO\,274-G001 (self-absorbing)]{Decomposition results for ESO\,274-G001 assuming a self-absorption model. Colour ranges have been chosen such that they contain 68\%, 95\% and 99.7\% of the distribution. Assuming a normal distribution, this represents the 1, 2 and 3$\sigma$ dispersion from the mean. Green lines are for the left side of the galaxy, cyan lines for the right side.  The grey band is the combined result.}
   \label{fig:ESO274-G001-velocitydispersion-thick}
\end{figure*}

\begin{figure*}
   \includegraphics[width=0.32\textwidth]{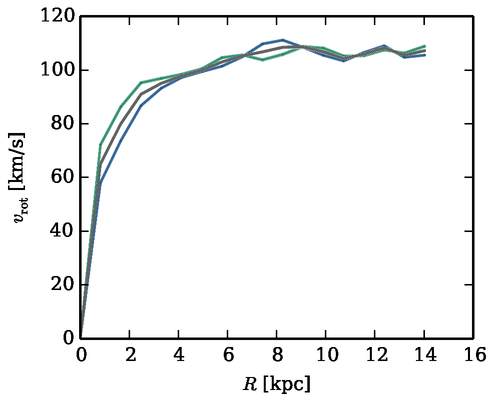}
   \includegraphics[width=0.32\textwidth]{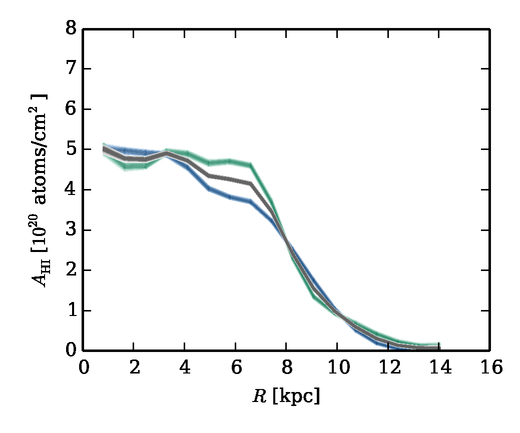}

   \includegraphics[width=0.32\textwidth]{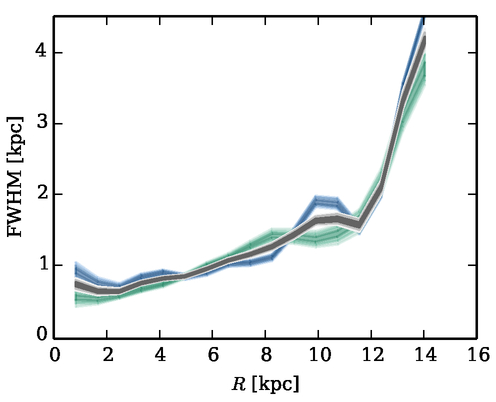}
   \includegraphics[width=0.32\textwidth]{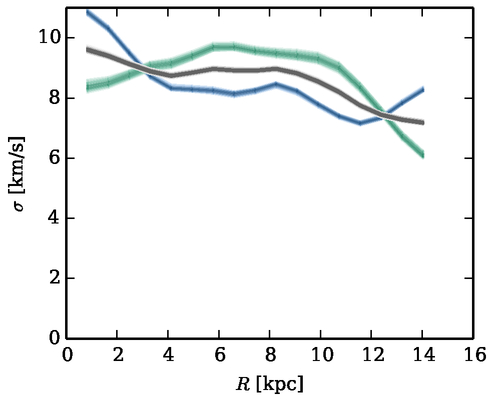}
\caption[\hi decomposition of UGC\,7321 (optically thin)]{Decomposition results for UGC\,7321 assuming an optically thin model. Colour ranges have been chosen such that they contain 68\%, 95\% and 99.7\% of the distribution. Assuming a normal distribution, this represents the 1, 2 and 3$\sigma$ dispersion from the mean. Green lines are for the left side of the galaxy, cyan lines for the right side.  The grey band is the combined result.}
   \label{fig:UGC7321-velocitydispersion-thin}
\end{figure*}

\begin{figure*}
   \includegraphics[width=0.32\textwidth]{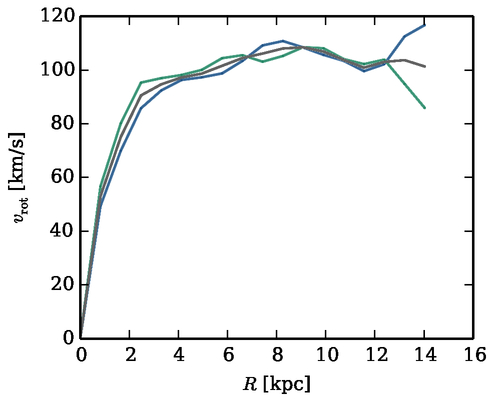}
   \includegraphics[width=0.32\textwidth]{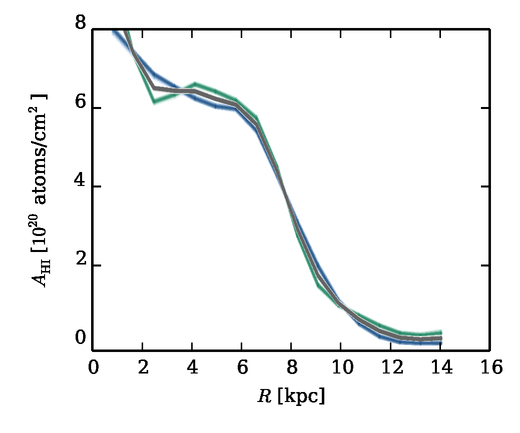}

   \includegraphics[width=0.32\textwidth]{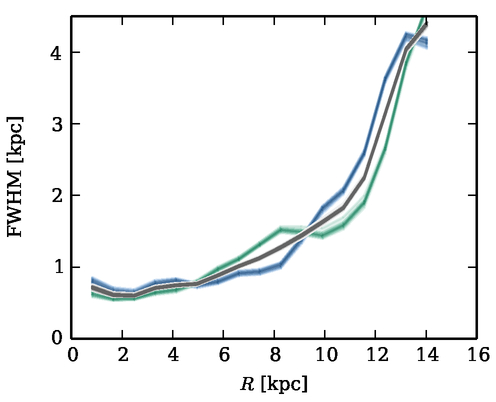}
   \includegraphics[width=0.32\textwidth]{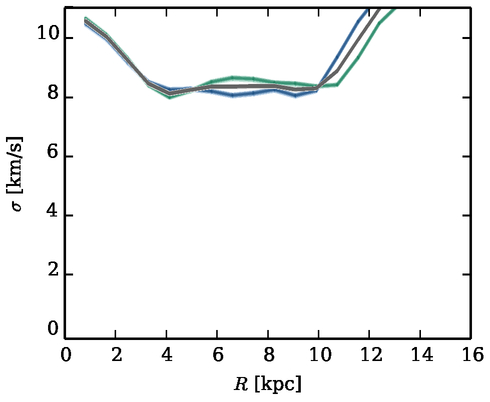}
\caption[\hi decomposition of UGC\,7321 (self-absorbing)]{Decomposition results for UGC\,7321 assuming a self-absorption model. Colour ranges have been chosen such that they contain 68\%, 95\% and 99.7\% of the distribution. Assuming a normal distribution, this represents the 1, 2 and 3$\sigma$ dispersion from the mean. Green lines are for the left side of the galaxy, cyan lines for the right side.  The grey band is the combined result.}
   \label{fig:UGC7321-velocitydispersion-thick}
\end{figure*}

\subsection{ESO\,274-G001}\label{sec:HIdecomposition_ESO274}
As we discussed in Section 5.7 in Paper I, galaxy ESO\,274-G001 
has a strong continuum source at its central position.
This causes strong \hi absorption, leading to our inability to  fit the 
central radii of the galaxy accurately.
We trust our results beyond 1.5\,kpc.
We show our optically thin and self-absorption fits in Figures 
\ref{fig:ESO274-G001-velocitydispersion-thin} and 
\ref{fig:ESO274-G001-velocitydispersion-thick}.

The rotation curve of the galaxy starts near 30\,km/s in the inner 
parts and rises continuously towards a maximum of 80\,km/s at 5\,kpc.
A similar result was found by \citet{OBrien2010C}.

The face-on surface density profile of the optically thin model 
hovers between 3 and $5\times10^{20}$ atoms/cm$^2$, with an average 
around $3.7\times10^{20}$ atoms/cm$^2$.
The face-on surface density profile of the self-absorption model 
has more \hi, in a plateau around $5\times10^{20}$ atoms/cm$^2$ up to 5\,kpc.
In both models, the \hi density drops rapidly towards zero at 8\,kpc.
The (optically thin) face-on surface density model of \citet{OBrien2010C} 
yielded a similar result as our optically thin model.

With the exception of the inner part, the thickness of 
the optically thin galaxy starts at 600\,pc at 2\,kpc radius, 
and increases to 750\,pc at a radius of 6\,kpc.
Similar to previous galaxies, the self-absorption yields a less 
thick disc, starting at 550\,pc at 2\,kpc, but only increasing to 
600\,pc at 6\,kpc radius.
Beyond 6\,kpc, both models yield a very large thickness, although 
the \hi has dropped to almost zero at those radii.
The flaring is very different from that measured by \citet{OBrien2010C}, 
whom found 100\,pc near the inner parts and flaring out linearly towards 
1.8\,kpc in the outer parts.

Both our models find a velocity dispersion that is dropping with radius. 
The optically thin model starts near 13\,km/s and drops towards 7\,km/s 
at 6\,kpc.
The self-absorption model drops towards 6\,km/s at that radius.
Beyond this radius, the velocity dispersion is unreliable. 
The velocity dispersion measured by \citet{OBrien2010C} shows a very 
different behavior, staying constant around 6.5\,km/s up to 6\,kpc.

The optically thin model has a total \hi mass of 
$3.1\pm0.02\times10^8$\,M$_\odot$.
The self-absorption model has a total mass of $4.1\pm0.1\times10^8$\,M$_\odot$.
A total of $25\pm1$\% of the \hi is thus hidden by self-absorption.

\subsection{UGC\,7321}
The \hi of galaxy UGC\,7321 has been previously modelled  by 
\citet{Uson2003A} and \citet{OBrien2010A}.
\citet{Matthews2003A} investigated the high-latitude \hi of 
the galaxy is detail.
The galaxy is the only in our sample which has been observed 
using the VLA.
Together with ESO\,274-G001, the galaxy has the highest signal 
to noise in our sample.
We show the results for the fits in Figures 
\ref{fig:UGC7321-velocitydispersion-thin} and 
\ref{fig:UGC7321-velocitydispersion-thick}.

The rotation curve rises steeply in the inner 
parts, only to level of around 9\,kpc at 110\,km/s. 
A similar behavior was reported by \citet{Uson2003A} and \citet{OBrien2010A}.

The thickness of the disc starts, in the optically 
thin model, near 750\,pc and increases linearly 
towards 1.5\,kpc at a radius of 10\,kpc. 
Beyond this radius, the thickness increases rapidly, 
due to the presence of a warp.
\citet{Uson2003A} estimate the onset of the warp at 
150'' (7.2\,kpc), although it becomes stronger beyond this radius.
Our self-absorption fit shows a different profile than 
the optically thin model. 
Between zero and five kpc, the thickness is nearly constant 
at 600-700\,pc, and only then flares out to 1.5\,kpc at 10\,kpc radius.
In Paper V we will only use the flaring 
out to 7.5 kpc for the fits to determine the axis ratio of the dark matter 
halo.

Both the optically thin model and the self-absorption 
model find a plateau in the face-on surface density profile.
This plateau remains constant out to roughly 7\,kpc, 
between 4 and $5\times10^{20}$\,atoms/cm$^2$ in the optically 
thin model, and 6 and $7\times10^{20}$\,atoms/cm$^2$ in the 
self-absorption model.
The optically thin results are consistent with both \citet{Uson2003A} and 
\citet{OBrien2010A}.

The high quality of the data has also allowed us to fit the velocity 
dispersion. 
The optically thin model hovers between 9 and 10\,km/s up to 10\,kpc radius.
The self-absorption model yields a decrease from 10 to 8\,km/s in the 
first 3\,kpc, but remains flat at 8\,km/s up to 10\,kpc radius.

The optically thin model has a total mass of $8.9\pm0.4\times10^8$\,M$_\odot$, 
while the self-absorption model has a total mass of 
$1.17\pm0.02\times10^9$\,M$_\odot$.
A total of $24\pm1$\% of the \hi is thus hidden by self-absorption.

\begin{figure*}[t]
   \includegraphics[width=0.32\textwidth]{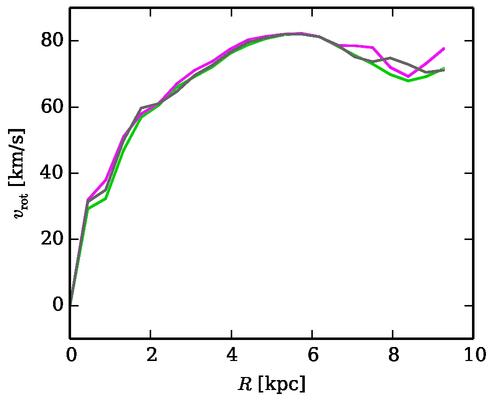}
   \includegraphics[width=0.32\textwidth]{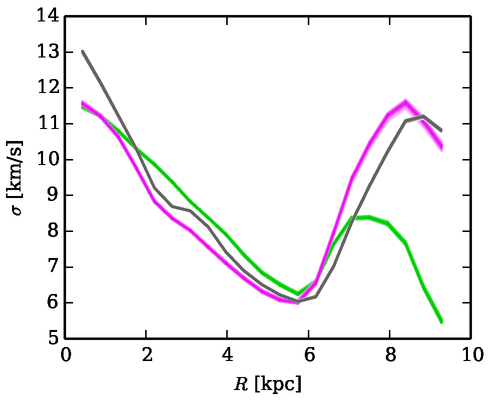}
\caption[Rotation curve and velocity dispersion with height]{Fits to the rotation curve and velocity dispersion as function of height. Measurement in ESO\,274-G001. The grey curve is the fit to the total galaxy, the cyan is the fit to the central plane of the galaxy and the green curve is the fit to the data that excludes the central plane.}\label{fig:sigmatest}
\end{figure*}

\subsection{Variations of the Velocity Dispersion with Height}\label{sec:velocitydispersionwithheight}
One of the (many) assumptions required for modelling the 
hydrostatics in a galaxy, is that the \hi velocity dispersion 
is constant with height (isothermal) and only varies with radius.
This has been assumed for a very long time, as for example by 
\citet{vdk81c}, based on the Galactic work done by \citet{Celnik1979A}, 
who found little evidence for variations.
To conclude this paper, we test this assumption.
We use the highly resolved observation of ESO\,274-G001.
We begin with the self-absorption results from Section 
\ref{sec:HIdecomposition_ESO274} and fix the flaring at the measured values.
A mask was created on the central  plane of the galaxy with a 
thickness of 290\,pc.
Two models were created, one in which only this central plane is 
visible, and one in which this plane is masked.
The results are shown in Figure \ref{fig:sigmatest}. 
The rotation curve of the high latitude fit can be seen to lag 
compared to the combined and main profiles.
The velocity dispersion of the high latitude fit is 1\,km/s 
larger than the central plane fit, with the combined fit falling 
in between the other two curves.

Note that this test is stretching the data to its absolute limits.
The error-bars are large; the only consolation is that we observe 
this behavior in the fits to the individual sides.
These results do not rule out that the \hi in galaxies is not isothermal.
It is beyond the scope of this project and the available data to 
test this in more detail in this series of papers.
It will be very interesting to explore this using the next 
generation of radio synthesis telescopes.

\section{Discussion \& Conclusions}
In the previous section, we have measured the structure and kinematics of 
six edge-on galaxies.
For each of these galaxies, we presented a fit that assumed an optically 
thin medium, as well as fit of a self-absorption medium with a spin 
temperature of 100\,K.
As already predicted in Section \ref{sec:whenmodelgobad}, not 
accounting for self-absorption leads to very different results 
than the self-absorption models give.
The most striking of this is the face-on column density, which 
is often measured 25\% lower in the inner parts of the optically 
thin models. 
Other problems that occur when assuming an optically thin medium 
are an overestimation of the thickness of the \hi disc and a 
completely wrong velocity dispersion.
In our test models (Section \ref{sec:whenmodelgobad}), the 
velocity dispersion tends to be too high in the inner parts and 
too low in the outer part, when fitting the optically thin results 
to a self-absorption model, although this exact behaviour might not 
be universal.

The implication of this work is that the \hi mass, thickness and 
velocity dispersion of galaxies are incorrect in the literature.

Comparing the optically thin results to those published in the literature, 
we find that our optically thin result match well.
Only the velocity dispersion of the optically thin model appear very 
different compared to the work by \citet{OBrien2010C}.
This difference is not fully understood, but is most likely due to 
the $z$-integrated PV diagram fitting by \citet{OBrien2010C}.
Self-absorption has the effect of flattening Gaussian profile of 
the \hi at a particular radius.
As the data is integrated with height, it will lead to a different 
observed slope at the outer edges, than would appear in a normal 
Gaussian profile.

\begin{table*}
\centering
\begin{tabular}{l|rr}
 Name & Optically thin mass [M$_\odot$] & Self-absorbing mass [M$_\odot$]\\\hline\hline
 IC\,5052&$7.4\pm0.7\times10^8$&$9.5\pm0.9\times10^8$\\
 IC\,5249&$4.8\pm0.2\times10^9$& $7.8\pm0.8\times10^9$\\
 ESO\,115-G021&$5.6\pm0.2\times10^9$&$7.2\pm0.1\times10^9$\\
 ESO\,138-G014&$3.4\pm0.1\times10^9$&$4.6\pm0.2\times10^9$\\
 ESO\,274-G001&$3.1\pm0.02\times10^8$&$4.1\pm0.1\times10^8$\\
 UGC\,7321&$8.9\pm0.4\times10^8$&$1.17\pm0.02\times10^9$\\\hline\hline
\end{tabular}
\caption[Total \hi mass of galaxies]{The total \hi mass of each galaxy as measured with our outer envelope strategy assuming either an optically thin or a self-absorbing mass.}
\label{table:HImassmeasured}
\end{table*}

In Table \ref{table:HImassmeasured}, we show the list of all \hi 
masses as measured using our outer envelope strategy.
On average we find that $27\pm6$\% of the \hi mass is hidden by 
self-absorption.
In Section 7 of Paper II, we predicted that 
self-absorption would hide 30\% of the mass if galaxy NGC\,2403 
had been viewed edge-on.
These results are thus compatible.
However, we stress that our analysis is based on a couple of 
important assumption.
The first is the constant spin temperature of 100\,K. 
While this appears to work well for the data, its value is 
chosen only for convenience.
In reality, the spin temperature is not constant, but is 
dependent on location inside the galaxy and the state of the \hi gas.
Assuming a  lower average spin temperature would have 
resulted in a much higher total \hi mass.

The second assumption is our treatment of the gas as a 
uniform medium.
As we already discussed in Section 2.3 of Paper II, 
this is not realistic. 
In our Galaxy, many lines of sight are known to be optically 
thick within a couple of hundred parsec, with the highest 
concentrations of \hi forming into distinct cloud structures 
\citep{Taylor2003,Allen2012A}.
The simulation of the Galactic plane by \citet{Douglas2010} 
shows that the opacity $\tau_\nu$ can go well above 25.
Due to beam smearing and the large distances, we cannot resolve those 
cloud structures in our galaxies and as such, a uniform medium is justified.
We stress however that because of the cloud structure of the \hi, it 
could well hiding even more \hi in the densest parts of these clouds.
An analysis of the cloud structure in the  nearby \emph{face-on} 
galaxy M\,31 by \citet{Braun2012A}, showed that $34\%$ of the \hi was
 already hidden. 
The \emph{edge-on} galaxies analyzed here could thus hide far more \hi 
than was measured here.

With these assumptions in mind, we conclude that our new fitting 
strategy, along with the presented self-absorption models, are a 
more accurate representation of the neutral hydrogen content of 
edge-on galaxies, than the optically-thin decomposition strategies 
presented here and by other authors.

\section*{Acknowledgments}
SPCP is grateful to the Space Telescope Science Institute, Baltimore, USA, the 
Research School for Astronomy and Astrophysics, Australian National University, 
Canberra, Australia, and the Instituto de Astrofisica de Canarias, La Laguna, 
Tenerife, Spain, for hospitality and support during  short and extended
working visits in the course of his PhD thesis research. He thanks
Roelof de Jong and Ron Allen for help and support during an earlier 
period as visiting student at Johns Hopkins University and 
the Physics and Astronomy Department, Krieger School of Arts and Sciences 
for this appointment.

PCK thanks the directors of these same institutions and his local hosts
Ron Allen, Ken Freeman and Johan Knapen for hospitality and support
during many work visits over the years, of which most were 
directly or indirectly related to the research presented in this series op 
papers.

Work visits by SPCP and PCK have been supported by an annual grant 
from the Faculty of Mathematics and Natural Sciences of 
the University of Groningen to PCK accompanying of his distinguished Jacobus 
C. Kapteyn professorhip and by the Leids Kerkhoven-Bosscha Fonds. PCK's work
visits were also supported by an annual grant from the Area  of Exact 
Sciences of the Netherlands Organisation for Scientific Research (NWO) in 
compensation for his membership of its Board.

%\begin{thebibliography}{99}

%\setlength{\bibsep}{0.1em}
\bibliography{refsIII}
\bibliographystyle{mn2e}

%\end{thebibliography}

%\onecolumn

%\appendix

%\section[]{}

\bsp

\label{lastpage}

\end{document}